\documentclass[a4paper,11pt]{article}
\usepackage{jheppub} 
\usepackage{adjustbox}
\usepackage{tikz}

\newcommand{\BGUE}{\text{BGUE}}
\newcommand{\Haar}{\text{Haar}}
\newcommand{\E}{\mathop{\mathbb{E}}}
\newcommand{\mE}{\mathcal{E}}
\newcommand{\one}{\bar{1}}
\newcommand{\two}{\bar{2}}
\newcommand{\tr}{\operatorname{tr}}

\newcommand{\mU}{\mathcal{U}}

\usepackage{tikz}
\newcommand*\circled[1]{\tikz[baseline=(char.base)]{\node[shape=circle,draw,inner sep=0.63pt] (char) {\scriptsize #1};}}

\title{\boldmath Brownian Gaussian Unitary Ensemble: non-equilibrium dynamics, efficient $k$-design and application in classical shadow tomography}

\author[a,b]{Haifeng Tang}
\affiliation[a]{Department of Physics, Stanford University, Stanford, California 94305, USA}
\affiliation[b]{Stanford Institute for Theoretical Physics, Stanford University, Stanford, California 94305, U.S.A.}

\emailAdd{hftang@stanford.edu}

\abstract{We construct and extensively study a Brownian generalization of the Gaussian Unitary Ensemble (BGUE). Our analysis begins with the non-equilibrium dynamics of BGUE, where we derive explicit analytical expressions for various one-replica and two-replica variables at finite \(N\) and \(t\). These variables include the spectral form factor and its fluctuation, the two-point function and its fluctuation, out-of-time-order correlators (OTOC), the second Rényi entropy, and the frame potential for unitary 2-designs. We discuss the implications of these results for hyperfast scrambling, emergence of tomperature, and replica-wormhole-like contributions in BGUE. Next, we investigate the low-energy physics of the effective Hamiltonian for an arbitrarily number of replicas, deriving long-time results for the frame potential. We conclude that the time required for the BGUE ensemble to reach \(k\)-design is linear in \(k\), consistent with previous findings in Brownian SYK models. Finally, we apply the BGUE model to the task of classical shadow tomography, deriving analytical results for the shadow norm and identifying an optimal time that minimizes the shadow norm, analogous to the optimal circuit depth in shallow-circuit shadow tomography.
}


\begin{document}
\maketitle
\flushbottom

\section{Introduction}

\paragraph{Motivation}
In recent work by Chen et al.~\cite{Chen:2024lxk}, an efficient approximation of a $k$-design of the Haar random unitary ensemble was explored using an $N \times N$ Hamiltonian $H$ drawn from the Gaussian Unitary Ensemble (GUE)~\cite{mehta2004random}. Due to energy conservation, $U_t = e^{iHt}$ cannot approach a $k$-design until a parametrically long time, which scales with some power of $N$~\cite{Cotler2017ChaosComplexityRMT, Roberts2017ChaosByDesign}. To overcome this obstacle, the authors proposed splitting the evolution into two ranges, where the system evolves under different GUE Hamiltonians in each range. Specifically, they considered the construction $U_t = e^{iH_1t/2} e^{iH_2t/2}$, where $H_1$ and $H_2$ are independently drawn from GUE. A key feature of this approach is that $H_1$ and $H_2$ generally do not commute, causing the spectrum of $U_t$ to mix quickly and distribute uniformly on the unit circle. This suggests that $e^{iH_1t/2} e^{iH_2t/2}$ might approach the Haar measure much faster on a timescale independent of $N$~\cite{Chen:2024lxk, Chen:2024ngb} (though dependent on $k$).

We extend this idea by further dividing $t$ into $n$ ranges, with each range evolving under a Hamiltonian $H_n$ independently drawn from GUE, forming $U_t = e^{iH_1t/n} e^{iH_2t/n} \cdots e^{iH_nt/n}$. As $n \rightarrow \infty$, this approach effectively creates a Brownian model, which we refer to as the Brownian Gaussian Unitary Ensemble (BGUE).

In this paper, we explore BGUE in detail. We first investigate several exactly solvable results for its non-equilibrium dynamics at finite $N$ and $t$. We then determine the time required for BGUE to approximate the Haar random ensemble, measured by $k$-design. Finally, we apply the BGUE model to classical shadow tomography, providing analytical results for the shadow norm and identifying an optimal time that minimizes it, akin to the optimal circuit depth in shallow-circuit shadow tomography schemes.

\paragraph{Model Considered}
We consider an $N \times N$ random Brownian Hamiltonian $H(t)$, where the matrix elements are zero-mean Gaussian random variables with the covariance:
\begin{equation}
\E \left[ H_{ij}(t) H_{kl}(t') \right] = N^{-1} \delta_{il} \delta_{jk} \delta(t-t')
\label{eq:definition of BGUE}
\end{equation}
All higher-order correlators satisfy Wick's theorem. The evolution operator is defined as $U_t = \mathcal{P}_{\text{path}} \left[ e^{i \int_0^t dt' H(t')} \right]$.

\paragraph{Summary of Results}
In section~\ref{section:1/2 replica}, we explore the non-equilibrium properties of BGUE. We calculate $\E [U_t \otimes U_t^*]$ and $\E [U_t \otimes U_t^* \otimes U_t \otimes U_t^*]$ exactly, providing explicit analytical expressions for various one-replica and two-replica variables as functions of arbitrary $N$ and $t$. These variables include the spectral form factor (SFF) (\ref{eq:SFF}), two-point functions (\ref{eq:2pt}), fluctuations of SFF (\ref{eq:variance of SFF}), fluctuations of two-point functions (\ref{eq:variance of 2pt}), out-of-time-order correlators (OTOC) (\ref{eq:OTOC}), the second Rényi entropy (\ref{eq:renyi entropy}), and the frame potential for unitary 2-designs (\ref{eq:frame potential}).

Our findings reveal that BGUE exhibits hyperfast scrambling and the emergence of tomperature~\cite{SusskindLin2022nss}. We also discuss the replica-wormhole-like contribution in BGUE, leading to a non-decaying two-point function as $t\rightarrow+\infty$~\cite{Saad:2019pqd, JuanMaldacena_2003} and non-vanishing fluctuations of the two-point function with a zero mean value~\cite{Stanford:2020wkf}.

In section~\ref{section:approaching k-design}, we examine the complexity of BGUE, determining the time required for BGUE to approximate the Haar ensemble in terms of $k$-design~\cite{dankert2005efficient, PhysRevA80012304, Gross_2007, 4262758, Cotler2017ChaosComplexityRMT, Roberts2017ChaosByDesign}. In section~\ref{section:low energy spectrum}, we study the low-energy eigen-wavefunctions, spectrum, and degeneracy of the effective imaginary-time evolution operator on $2k$-replicated contours. Based on these results, in section~\ref{section:calculating frame potential}, we obtain the frame potential $F^{(k)}_\BGUE(t)$ for large but finite $t$ (\ref{eq:frame potential for k-replica}) for arbitrary $k$, and the corresponding time (\ref{eq:time to approach k-design}) needed to approach a unitary $k$-design. This timescale is linear in $k \log N$, consistent with previous studies on the Brownian Sachdev-Ye-Kitaev (SYK) model~\cite{Sachdev2015, Maldecena2016, Jian:2022pvj}. In appendix~\ref{section: subset of high energy spectrum}, we derive a subset of the high-energy spectrum.

In section~\ref{section: applications to shadow tomography}, we apply BGUE to classical shadow tomography~\cite{Huang2020}, obtaining analytical results for the shadow norm (\ref{eq: shadow norm}) and identifying an optimal time (\ref{eq:optimal time of shadow norm}) when the observable $O$ includes both diagonal and off-diagonal parts. This optimal time parallels the optimal circuit depth in shallow-circuit shadow tomography~\cite{PhysRevLett.130.230403, PhysRevResearch.5.023027}.

\paragraph{Note Added.}
After completing this work, we noted the appearance of reference~\cite{Guo:2024zmr} on arXiv, which also studies the complexity of the BGUE model in their section 4. While we agree on results regarding the frame potential and $k$-design using different approaches, reference~\cite{Guo:2024zmr} focuses on the concept and formalism of complexity in more generic Brownian models, including the Brownian Sachdev-Ye-Kitaev (SYK) model. Our work emphasizes on BGUE itself, examining many non-equilibrium properties and its applications in some quantum information tasks.

\section{Non-equilibrium dynamics: exact results for one/two-replica variables}
\label{section:1/2 replica}

In this section, we investigate various aspects of the non-equilibrium dynamics of BGUE. For spectral properties, we study the Spectral Form Factor (SFF) and its fluctuation~\cite{Cotler2017BHandRMT}. For correlation functions and scrambling behavior~\cite{Maldacena2016bound,YasuhiroSekino2008,Maldecena2016}, we examine the two-point function and its fluctuation, as well as OTOC~\cite{Maldecena2016,Roberts2018,Qi2019,Lucas2019}. Additionally, we study the second Rényi entropy of pure product states evolved under BGUE as a measure of entanglement generation~\cite{RQC1,RQC2,RQC3}.

These quantities depend on one or two replicas of $U_t \otimes U_t^*$; therefore, it is beneficial to first derive the averaged evolution operators on two/four contours: $\mathcal{U}_1(t) \equiv \E[U_t \otimes U_t^*]$ and $\mathcal{U}_2(t) \equiv \E[U_t \otimes U_t^* \otimes U_t \otimes U_t^*]$. This calculation scheme is similar to random tensor network models~\cite{RQC5} and random unitary circuit models~\cite{RQC1,RQC2,RQC3,RQC4,RQC6} extensively studied in previous literature concerning scrambling dynamics~\cite{RQC1,RQC2}, Measurement Induced Phase Transition (MIPT)~\cite{koh2022experimental,li2022cross,MIPT1,MIPT10,MIPT2,MIPT24,MIPT3,MIPT30,MIPT5,MIPT7,noel2022measurement}, and the holographic theory of gravity~\cite{Milekhin:2023bjv,Stanford:2023npy,Open1}. A common feature of these models is that, due to the Brownian property, Hamiltonians (for continuous time evolution) or unitaries (for discrete time evolution) at different times are uncorrelated, meaning that the average taken at different times factorizes. For generic $k$-replica observables in BGUE, we ultimately obtain an imaginary time evolution on $2k$-contours: 
\begin{equation}
\mathcal{U}_k(t) \equiv \E[U_t^{\otimes k} \otimes U_t^{*\otimes k}] \equiv e^{\mathcal{L}_k t}
\label{eq:U_k(t)=exp L_k t}
\end{equation}
where $\mathcal{L}_k$ is a Hermitian operator acting on $2k$-contours. We make four further comments that might help readers better interpret equation~(\ref{eq:U_k(t)=exp L_k t}):
\begin{itemize}
\item[1.]\textit{Time translation invariance of $\mathcal{L}_k$:} Although a single realization of BGUE lacks time translation invariance, it is restored after averaging.
\item[2.]\textit{$\mathcal{U}_k(t)$ satisfies an ODE:} Namely $\partial_t \mathcal{U}_k(t) = \mathcal{L}_k \mathcal{U}_k(t)$. This is due to the Markovian property of Brownian models. For a static random Hamiltonian like GUE or SYK-model~\cite{Maldecena2016}, we expect a more complicated differential-integral equation (the Schwinger-Dyson equation for SYK).
\item[3.]\textit{$\mathcal{U}_k(t)$ is not unitary:} This is because a linear superposition of unitary matrices is generally not unitary.
\item[4.]\textit{$\mathcal{L}_k$ is Hermitian:} This follows from two reasons: the time-translation symmetry of the ensemble and the zero mean of $H(t)$, where the latter implies an additional symmetry $H(t) \rightarrow -H(t)$.
\end{itemize}

Prepared with this high-level outline of the calculation structure, we explicitly derive $\mathcal{L}_1, \mathcal{L}_2$ and $\mathcal{U}_1(t), \mathcal{U}_2(t)$ in section~\ref{section:Averaged evolution operator on one and two replica space}, and then apply these results to calculate observables in sections~\ref{section:1 replica observable} and~\ref{section:2 replica observable}.

\subsection{Averaged evolution operator on one and two replica space}
\label{section:Averaged evolution operator on one and two replica space}

We first need to set up some notation. For $k$-replica (equivalently $2k$-contour) variables, we use $i \in \{1,...,k\}$ to denote the $i^{\text{th}}$ forward contour and $\bar{j} \in \{1,...,k\}$ to denote the $j^{\text{th}}$ backward contour. Similarly, we use $H_i(t), H_{\bar{j}}(t)$ to denote the Hamiltonians on the corresponding contours. Using the definition of BGUE in equation~(\ref{eq:definition of BGUE}), we obtain three important results:
\begin{equation}
\begin{aligned}
&\E \left[H_i(t)^2\right]=\E \adjincludegraphics[valign=c, width=0.23\textwidth]{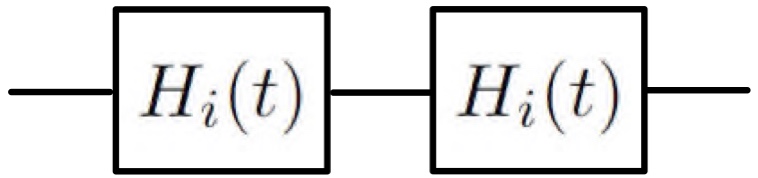}=\Delta t^{-1}N^{-1}\adjincludegraphics[valign=c, width=0.17\textwidth]{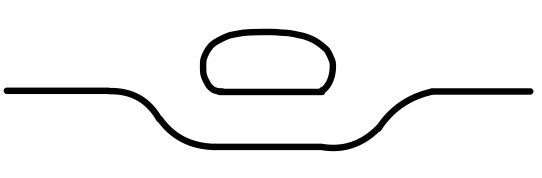}=\Delta t^{-1}\adjincludegraphics[valign=c, width=0.1\textwidth]{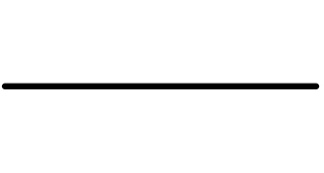}=\Delta t^{-1}\\
&\E \left[H_{\bar{i}}^*(t)^2\right]=\E \adjincludegraphics[valign=c, width=0.227\textwidth]{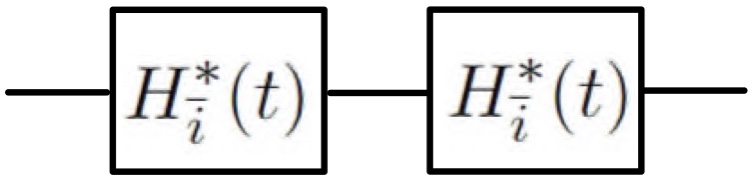}=\Delta t^{-1}N^{-1}\adjincludegraphics[valign=c, width=0.17\textwidth]{HH_3.jpg}=\Delta t^{-1}\adjincludegraphics[valign=c, width=0.1\textwidth]{HH_4.jpg}=\Delta t^{-1}\\
\end{aligned}
\label{eq: HH}
\end{equation}
\begin{equation}
\begin{aligned}
&\E\left[H_i(t)\otimes H_j(t)\right]=\E \adjincludegraphics[valign=c, width=0.13\textwidth]{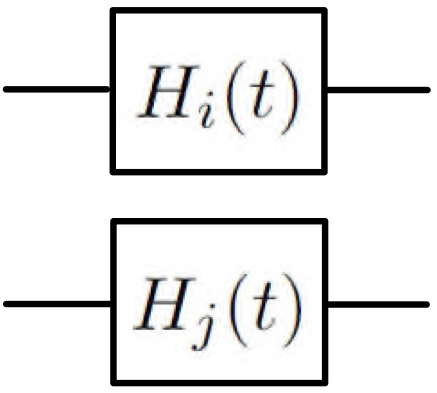}=\Delta t^{-1}N^{-1}\adjincludegraphics[valign=c, width=0.1\textwidth]{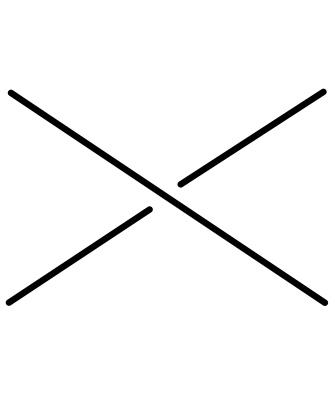}\equiv\Delta t^{-1}N^{-1}X_{ij},\ i\neq j\\
&\E\left[H_{\bar{i}}^*(t)\otimes H_{\bar{j}}^*(t)\right]=\E \adjincludegraphics[valign=c, width=0.13\textwidth]{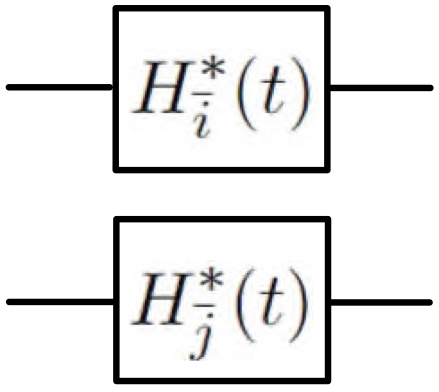}=\Delta t^{-1}N^{-1}\adjincludegraphics[valign=c, width=0.1\textwidth]{HH_7.jpg}\equiv\Delta t^{-1}N^{-1}X_{\bar{i}\bar{j}},\ \bar{i}\neq \bar{j}\\
\end{aligned}
\label{eq:define X_ij}
\end{equation}
\begin{equation}
\E\left[H_i(t)\otimes H_{\bar{j}}^*(t)\right]=\E \adjincludegraphics[valign=c, width=0.13\textwidth]{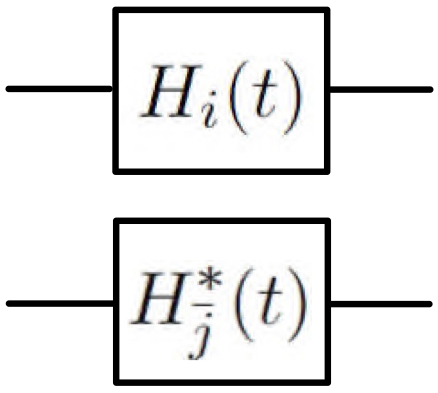}=\Delta t^{-1}N^{-1}\adjincludegraphics[valign=c, width=0.1\textwidth]{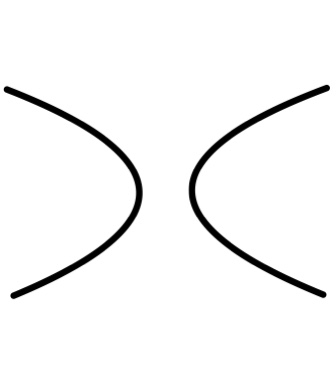}\equiv\Delta t^{-1}N^{-1}P_{i\bar{j}}
\label{eq: define P_ij}
\end{equation}
where $\Delta t$ is defined through the regularization of $\delta(t-t')\approx \frac{\delta_{t,t'}}{\Delta t}$ (or equivalently interpreted as Trotterization of continuous time into discrete time). Here, $X_{ij}$ is the SWAP operator between the $i,j$ contours, and $P_{i\bar{j}}\equiv|\text{EPR}_{i\bar{j}}\rangle\langle\text{EPR}_{i\bar{j}}|$ is the unnormalized projector (we mean that the state $|\text{EPR}_{i\bar{j}}\rangle$ is unnormalized: $\langle\text{EPR}_{i\bar{j}}|\text{EPR}_{i\bar{j}}\rangle=N$) onto the maximally entangled state linking $i$ and $\bar{j}$.

Armed with equations~(\ref{eq: HH}),~(\ref{eq:define X_ij}), and~(\ref{eq: define P_ij}), we are now ready to calculate $\mathcal{L}_k$:
\begin{equation}
\begin{aligned}
\lim_{\Delta t\rightarrow 0}\E\left[\mathcal{U}_k(\Delta t)\right]&=\E\left[\exp\left(i\Delta t\sum_{i=1}^k H_i(t)-i\Delta t\sum_{\bar{j}=1}^k H_{\bar{j}}^*(t)\right)\right]\\
&=1+\Delta t^2\Bigg(-\frac{1}{2}\sum_{i=1}^k\E[H_i(t)^2]-\frac{1}{2}\sum_{\bar{j}=1}^k\E[H_{\bar{j}}^*(t)^2]+\sum_{i,\bar{j}=1}^k\E[H_i(t)\otimes H_{\bar{j}}^*(t)]\\
&\ \ \ \ \ \ \ \ \ \ \ \ \ \ \ \ \ -\sum_{1\leq i<j\leq k}\E[H_i(t)\otimes H_j(t)]-\sum_{1\leq \bar{i}<\bar{j}\leq k}\E[H_{\bar{i}}^*(t)\otimes H_{\bar{j}}^*(t)]\Bigg)+O(\Delta t^4)\\
&=1+\Delta t\Bigg(-k+N^{-1}\sum_{i,\bar{j}=1}^kP_{i\bar{j}}-N^{-1}\sum_{1\leq i<j\leq k}X_{ij}-N^{-1}\sum_{1\leq \bar{i}<\bar{j}\leq k}X_{\bar{i}\bar{j}}\Bigg)+O(\Delta t^2)\\
&\equiv 1+\Delta t\cdot\mathcal{L}_k+O(\Delta t^2)
\end{aligned}
\end{equation}
where we expand the exponential and drop the linear-in-$\Delta t$ term since $H(t)$ has zero mean. We find that:
\begin{equation}
\mathcal{L}_k=-k+N^{-1}\sum_{i,\bar{j}=1}^kP_{i\bar{j}}-N^{-1}\sum_{1\leq i<j\leq k}X_{ij}-N^{-1}\sum_{1\leq \bar{i}<\bar{j}\leq k}X_{\bar{i}\bar{j}}
\label{eq: L_k}
\end{equation}

Exponentiating $\mathcal{L}_k$ to obtain $\mathcal{U}_k(t)=e^{\mathcal{L}_k t}$ is challenging for generic $k$. This is because $\{P_{i\bar{j}},X_{ij},X_{\bar{i}\bar{j}}\}$ are the generators of a non-trivial algebra, which is a sub-algebra of the partition algebra~\cite{enwiki:1170404101}. Therefore, the task of exponentiating $\mathcal{L}_k$ is equivalent to finding the corresponding representation of this algebra.

Fortunately, for the special case of $k=1,2$, we can calculate $\mathcal{U}_k(t)$ exactly, using the idea of Krylov space described below.

\paragraph{Two-contour averaged evolution operator.} According to equation~(\ref{eq: L_k}), $\mathcal{L}_1$ is very simple:
\begin{equation}
\mathcal{L}_1=-1+N^{-1}P_{1\one}
\end{equation}
Using the property $P_{i\bar{j}}^2=N P_{i\bar{j}}$, we obtain:
\begin{equation}
\mathcal{U}_1(t)=e^{-t}\mathbb{I}+(1-e^{-t})N^{-1}P_{1\one}
\label{eq:U_1}
\end{equation}
A quick consistency check is to calculate $\tr[U_tU_t^{\dagger}]=N$ using $\mathcal{U}_1(t)$, which is realized by $\langle\text{EPR}_{1\one}|\mathcal{U}_1|\text{EPR}_{1\one}\rangle=e^{-t}N+(1-e^{-t})N=N$.

We also notice that when $t\rightarrow+\infty$, $\mathcal{U}_1(+\infty)=N^{-1}P_{1\one}$ matches the expectation from the Haar random ensemble. This suggests that our BGUE smoothly interpolates between the identity unitary and the Haar ensemble through time evolution. We will see more evidence of this in higher replicas.

\paragraph{Four-contour averaged evolution operator.} Similarly, using equation~(\ref{eq: L_k}), we obtain $\mathcal{L}_2$:
\begin{equation}
\mathcal{L}_2=-2+N^{-1}(P_{1\one}+P_{2\two}+P_{1\two}+P_{2\one})-N^{-1}(X_{12}+X_{\one\two})
\end{equation}
where $X_{12}$ is the SWAP operator between 1 and 2. 

Next, we need to exponentiate $\mathcal{L}_2$, which is more complicated than exponentiating $\mathcal{L}_1$. The trick is to notice that the Krylov space~\cite{Parker2019,Bhattacharyya2023,Erdmenger2023,Kar2022,Liu2023,Tang:2023ocr,Tang:2024xgg,Nandy:2024htc} of $\mathcal{L}_2$, defined as polynomials of $\mathcal{L}_2$, $\mathcal{K} \equiv \text{span}\{1,\mathcal{L}_2,\mathcal{L}_2^2,\mathcal{L}_2^3,\cdots\}$, is finite-dimensional. This is because the product of links are still links, and there are a total of $4!=24$ types of links, corresponding to the 24 permutation elements of $\mathbb{S}^4$. Further dimension reduction can be made by noting the symmetry of $\mathcal{L}_2$: permutation of four contours and Hermiticity. We group all 24 link operators into 8 categories, serving as the basis of the Krylov space, namely $\mathcal{K} = \text{span}\{A_i|i=1,\cdots,8\}$:

\begin{equation}
\begin{aligned}
&A_1=
\begin{array}{c}
\begin{tikzpicture}[scale=0.29]
    \foreach \y in {1,2,3,4} {
        \coordinate (l\y) at (0, -\y);
    }
    \foreach \y in {1,2,3,4} {
        \coordinate (r\y) at (3.5, -\y);
    }

    \draw[thick] (l1) -- (r1);
    \draw[thick] (l2) -- (r2);
    \draw[thick] (l3) --(r3);
    \draw[thick] (l4) --(r4);

    \node[text=gray] at (4,-1) {\tiny 1};
    \node[text=gray] at (4,-2) {\tiny $\bar{1}$};
    \node[text=gray] at (4,-3) {\tiny 2};
    \node[text=gray] at (4,-4) {\tiny $\bar{2}$};
\end{tikzpicture}
\end{array}=1\\
&A_2=
\begin{array}{c}
\begin{tikzpicture}[scale=0.29]
    \foreach \y in {1,2,3,4} {
        \coordinate (l\y) at (0, -\y);
    }
    \foreach \y in {1,2,3,4} {
        \coordinate (r\y) at (3.5, -\y);
    }

    \draw[thick] (l3) -- (r3);
    \draw[thick] (l4) -- (r4);
    \draw[thick] (l1) .. controls+(right:1cm) and +(right:1cm)..(l2);
    \draw[thick] (r1) .. controls +(left:1cm) and +(left:1cm) .. (r2);
\end{tikzpicture}
\end{array}+
\begin{array}{c}
\begin{tikzpicture}[scale=0.29]
    \foreach \y in {1,2,3,4} {
        \coordinate (l\y) at (0, -\y);
    }
    \foreach \y in {1,2,3,4} {
        \coordinate (r\y) at (3.5, -\y);
    }

    \draw[thick] (l1) -- (r1);
    \draw[thick] (l2) -- (r2);
    \draw[thick] (l3) .. controls+(right:1cm) and +(right:1cm)..(l4);
    \draw[thick] (r3) .. controls +(left:1cm) and +(left:1cm) .. (r4);
\end{tikzpicture}
\end{array}+
\begin{array}{c}
\begin{tikzpicture}[scale=0.29]
    \foreach \y in {1,2,3,4} {
        \coordinate (l\y) at (0, -\y);
    }
    \foreach \y in {1,2,3,4} {
        \coordinate (r\y) at (3.5, -\y);
    }

    \draw[thick] (l2) -- (r2);
    \draw[thick] (l3) -- (r3);
    \filldraw [white] (0.7,-2) circle (7pt);
    \filldraw [white] (0.7,-3) circle (7pt);
    \filldraw [white] (2.8,-2) circle (7pt);
    \filldraw [white] (2.8,-3) circle (7pt);
    \draw[thick] (l1) .. controls+(right:1cm) and +(right:1cm)..(l4);
    \draw[thick] (r1) .. controls +(left:1cm) and +(left:1cm) .. (r4);
\end{tikzpicture}
\end{array}+
\begin{array}{c}
\begin{tikzpicture}[scale=0.29]
    \foreach \y in {1,2,3,4} {
        \coordinate (l\y) at (0, -\y);
    }
    \foreach \y in {1,2,3,4} {
        \coordinate (r\y) at (3.5, -\y);
    }

    \draw[thick] (l1) -- (r1);
    \draw[thick] (l4) -- (r4);
    \draw[thick] (l2) .. controls+(right:1cm) and +(right:1cm)..(l3);
    \draw[thick] (r3) .. controls +(left:1cm) and +(left:1cm) .. (r2);
\end{tikzpicture}
\end{array}=P_{1\one}+P_{2\two}+P_{1\two}+P_{2\one}\\
&A_3=
\begin{array}{c}
\begin{tikzpicture}[scale=0.29]
    \foreach \y in {1,2,3,4} {
        \coordinate (l\y) at (0, -\y);
    }
    \foreach \y in {1,2,3,4} {
        \coordinate (r\y) at (3.5, -\y);
    }
    
    \draw[thick] (l2) -- (r2);
    \draw[thick] (l4) -- (r4);
    \filldraw [white] (1.75,-2) circle (13pt);
    \draw[thick] (l1) -- (r3);
    \draw[thick] (l3) -- (r1);

\end{tikzpicture}
\end{array}+
\begin{array}{c}
\begin{tikzpicture}[scale=0.29]
    \foreach \y in {1,2,3,4} {
        \coordinate (l\y) at (0, -\y);
    }
    \foreach \y in {1,2,3,4} {
        \coordinate (r\y) at (3.5, -\y);
    }

    \draw[thick] (l1) -- (r1);
    \draw[thick] (l3) -- (r3);
    \filldraw [white] (1.75,-3) circle (13pt);
    \draw[thick] (l2) -- (r4);
    \draw[thick] (l4) -- (r2);
\end{tikzpicture}
\end{array}=X_{12}+X_{\one\two}\\
&A_4=\begin{array}{c}
\begin{tikzpicture}[scale=0.29]
    \foreach \y in {1,2,3,4} {
        \coordinate (l\y) at (0, -\y);
    }
    \foreach \y in {1,2,3,4} {
        \coordinate (r\y) at (3.5, -\y);
    }

    \draw[thick] (l3) -- (r1);
    \draw[thick] (l4) -- (r4);
    \draw[thick] (l1) .. controls+(right:1cm) and +(right:1cm)..(l2);
    \draw[thick] (r2) .. controls +(left:1cm) and +(left:1cm) ..(r3);
\end{tikzpicture}
\end{array}+
\begin{array}{c}
\begin{tikzpicture}[scale=0.29]
    \foreach \y in {1,2,3,4} {
        \coordinate (l\y) at (0, -\y);
    }
    \foreach \y in {1,2,3,4} {
        \coordinate (r\y) at (3.5, -\y);
    }

    \draw[thick] (l1) -- (r3);
    \draw[thick] (l4) -- (r4);
    \draw[thick] (l3) .. controls+(right:1cm) and +(right:1cm)..(l2);
    \draw[thick] (r2) .. controls +(left:1cm) and +(left:1cm) ..(r1);
\end{tikzpicture}
\end{array}+
\begin{array}{c}
\begin{tikzpicture}[scale=0.29]
    \foreach \y in {1,2,3,4} {
        \coordinate (l\y) at (0, -\y);
    }
    \foreach \y in {1,2,3,4} {
        \coordinate (r\y) at (3.5, -\y);
    }

    \draw[thick] (l4) -- (r2);
    \draw[thick] (l1) -- (r1);
    \draw[thick] (l2) .. controls+(right:1cm) and +(right:1cm)..(l3);
    \draw[thick] (r3) .. controls +(left:1cm) and +(left:1cm) ..(r4);
\end{tikzpicture}
\end{array}+
\begin{array}{c}
\begin{tikzpicture}[scale=0.29]
    \foreach \y in {1,2,3,4} {
        \coordinate (l\y) at (0, -\y);
    }
    \foreach \y in {1,2,3,4} {
        \coordinate (r\y) at (3.5, -\y);
    }

    \draw[thick] (l2) -- (r4);
    \draw[thick] (l1) -- (r1);
    \draw[thick] (l3) .. controls+(right:1cm) and +(right:1cm)..(l4);
    \draw[thick] (r2) .. controls +(left:1cm) and +(left:1cm) ..(r3);
\end{tikzpicture}
\end{array}+
\begin{array}{c}
\begin{tikzpicture}[scale=0.29]
    \foreach \y in {1,2,3,4} {
        \coordinate (l\y) at (0, -\y);
    }
    \foreach \y in {1,2,3,4} {
        \coordinate (r\y) at (3.5, -\y);
    }
    \draw[thick] (l2) -- (r2);
    \filldraw [white] (1.75,-2) circle (10pt);
    \draw[thick] (l1) -- (r3);
    \filldraw [white] (2.8,-2) circle (6pt);
    \filldraw [white] (2.8,-2.6) circle (6pt);
    \draw[thick] (l3) .. controls+(right:1cm) and +(right:1cm)..(l4);
    \draw[thick] (r1) .. controls +(left:1cm) and +(left:1cm) ..(r4);
\end{tikzpicture}
\end{array}+
\begin{array}{c}
\begin{tikzpicture}[scale=0.29]
    \foreach \y in {1,2,3,4} {
        \coordinate (l\y) at (0, -\y);
    }
    \foreach \y in {1,2,3,4} {
        \coordinate (r\y) at (3.5, -\y);
    }
    \draw[thick] (l2) -- (r2);
    \filldraw [white] (1.75,-2) circle (10pt);
    \draw[thick] (l3) -- (r1);
    \filldraw [white] (0.7,-2) circle (6pt);
    \filldraw [white] (0.7,-2.6) circle (6pt);
    \draw[thick] (l1) .. controls+(right:1cm) and +(right:1cm)..(l4);
    \draw[thick] (r3) .. controls +(left:1cm) and +(left:1cm) ..(r4);
\end{tikzpicture}
\end{array}+
\begin{array}{c}
\begin{tikzpicture}[scale=0.29]
    \foreach \y in {1,2,3,4} {
        \coordinate (l\y) at (0, -\y);
    }
    \foreach \y in {1,2,3,4} {
        \coordinate (r\y) at (3.5, -\y);
    }
    \draw[thick] (l3) -- (r3);
    \filldraw [white] (1.75,-3) circle (10pt);
    \draw[thick] (l4) -- (r2);
    \filldraw [white] (2.8,-3) circle (6pt);
    \filldraw [white] (2.8,-2.4) circle (6pt);
    \draw[thick] (l1) .. controls+(right:1cm) and +(right:1cm)..(l2);
    \draw[thick] (r1) .. controls +(left:1cm) and +(left:1cm) ..(r4);
\end{tikzpicture}
\end{array}+
\begin{array}{c}
\begin{tikzpicture}[scale=0.29]
    \foreach \y in {1,2,3,4} {
        \coordinate (l\y) at (0, -\y);
    }
    \foreach \y in {1,2,3,4} {
        \coordinate (r\y) at (3.5, -\y);
    }

    \draw[thick] (l3) -- (r3);
    \filldraw [white] (1.75,-3) circle (10pt);
    \draw[thick] (l2) -- (r4);
    \filldraw [white] (0.7,-3) circle (6pt);
    \filldraw [white] (0.7,-2.4) circle (6pt);
    \draw[thick] (l1) .. controls+(right:1cm) and +(right:1cm)..(l4);
    \draw[thick] (r1) .. controls +(left:1cm) and +(left:1cm) ..(r2);
\end{tikzpicture}
\end{array}\\
&A_5=\begin{array}{c}
\begin{tikzpicture}[scale=0.29]
    \foreach \y in {1,2,3,4} {
        \coordinate (l\y) at (0, -\y);
    }
    \foreach \y in {1,2,3,4} {
        \coordinate (r\y) at (3.5, -\y);
    }

    \draw[thick] (l1) .. controls+(right:1cm) and +(right:1cm)..(l2);
    \draw[thick] (l3) .. controls+(right:1cm) and +(right:1cm)..(l4);
    \draw[thick] (r1) .. controls +(left:1cm) and +(left:1cm) ..(r2);
    \draw[thick] (r3) .. controls +(left:1cm) and +(left:1cm) ..(r4);
\end{tikzpicture}
\end{array}+
\begin{array}{c}
\begin{tikzpicture}[scale=0.29]
    \foreach \y in {1,2,3,4} {
        \coordinate (l\y) at (0, -\y);
    }
    \foreach \y in {1,2,3,4} {
        \coordinate (r\y) at (3.5, -\y);
    }

    \draw[thick] (l1) .. controls+(right:1.5cm) and +(right:1.5cm)..(l4);
    \draw[thick] (l3) .. controls+(right:1cm) and +(right:1cm)..(l2);
    \draw[thick] (r1) .. controls +(left:1.5cm) and +(left:1.5cm) ..(r4);
    \draw[thick] (r3) .. controls +(left:1cm) and +(left:1cm) ..(r2);
\end{tikzpicture}
\end{array}\\
&A_6=\begin{array}{c}
\begin{tikzpicture}[scale=0.29]
    \foreach \y in {1,2,3,4} {
        \coordinate (l\y) at (0, -\y);
    }
    \foreach \y in {1,2,3,4} {
        \coordinate (r\y) at (3.5, -\y);
    }

    \draw[thick] (l1) -- (r3);
    \draw[thick] (l2) -- (r4);
    \filldraw [white] (1.75,-2) circle (7pt);
    \filldraw [white] (0.875,-2.5) circle (7pt);
    \filldraw [white] (2.625,-2.5) circle (7pt);
    \filldraw [white] (1.75,-3) circle (7pt);
    \draw[thick] (l3) -- (r1);
    \draw[thick] (l4) -- (r2);
\end{tikzpicture}
\end{array}\\
&A_7=\begin{array}{c}
\begin{tikzpicture}[scale=0.29]
    \foreach \y in {1,2,3,4} {
        \coordinate (l\y) at (0, -\y);
    }
    \foreach \y in {1,2,3,4} {
        \coordinate (r\y) at (3.5, -\y);
    }

    \draw[thick] (l1) .. controls+(right:1cm) and +(right:1cm)..(l2);
    \draw[thick] (l3) .. controls+(right:1cm) and +(right:1cm)..(l4);
    \draw[thick] (r1) .. controls +(left:1.5cm) and +(left:1.5cm) ..(r4);
    \draw[thick] (r3) .. controls +(left:1cm) and +(left:1cm) ..(r2);
\end{tikzpicture}
\end{array}+
\begin{array}{c}
\begin{tikzpicture}[scale=0.29]
    \foreach \y in {1,2,3,4} {
        \coordinate (l\y) at (0, -\y);
    }
    \foreach \y in {1,2,3,4} {
        \coordinate (r\y) at (3.5, -\y);
    }

    \draw[thick] (l1) .. controls+(right:1.5cm) and +(right:1.5cm)..(l4);
    \draw[thick] (l3) .. controls+(right:1cm) and +(right:1cm)..(l2);
    \draw[thick] (r1) .. controls +(left:1cm) and +(left:1cm) ..(r2);
    \draw[thick] (r3) .. controls +(left:1cm) and +(left:1cm) ..(r4);
\end{tikzpicture}
\end{array}\\
&A_8=\begin{array}{c}
\begin{tikzpicture}[scale=0.29]
    \foreach \y in {1,2,3,4} {
        \coordinate (l\y) at (0, -\y);
    }
    \foreach \y in {1,2,3,4} {
        \coordinate (r\y) at (3.5, -\y);
    }

    \draw[thick] (l3) -- (r1);
    \draw[thick] (l4) -- (r2);
    \draw[thick] (l1) .. controls +(right:1cm) and +(right:1cm) ..(l2);
    \draw[thick] (r3) .. controls +(left:1cm) and +(left:1cm) ..(r4);
\end{tikzpicture}
\end{array}+
\begin{array}{c}
\begin{tikzpicture}[scale=0.29]
    \foreach \y in {1,2,3,4} {
        \coordinate (l\y) at (0, -\y);
    }
    \foreach \y in {1,2,3,4} {
        \coordinate (r\y) at (3.5, -\y);
    }

    \draw[thick] (l1) -- (r3);
    \draw[thick] (l2) -- (r4);
    \draw[thick] (l3) .. controls +(right:1cm) and +(right:1cm) ..(l4);
    \draw[thick] (r1) .. controls +(left:1cm) and +(left:1cm) ..(r2);
\end{tikzpicture}
\end{array}+
\begin{array}{c}
\begin{tikzpicture}[scale=0.29]
    \foreach \y in {1,2,3,4} {
        \coordinate (l\y) at (0, -\y);
    }
    \foreach \y in {1,2,3,4} {
        \coordinate (r\y) at (3.5, -\y);
    }

    \draw[thick] (l2) -- (r4);
    \filldraw [white] (0.875,-2.5) circle (6pt);
    \draw[thick] (l3) -- (r1);
    \filldraw [white] (1.5,-2.2) circle (5pt);
    \filldraw [white] (1.5,-2.8) circle (5pt);
    \draw[thick] (l1) .. controls +(right:2cm) and +(right:2cm) ..(l4);
    \draw[thick] (r3) .. controls +(left:1cm) and +(left:1cm) ..(r2);
\end{tikzpicture}
\end{array}+
\begin{array}{c}
\begin{tikzpicture}[scale=0.29]
    \foreach \y in {1,2,3,4} {
        \coordinate (l\y) at (0, -\y);
    }
    \foreach \y in {1,2,3,4} {
        \coordinate (r\y) at (3.5, -\y);
    }

    \draw[thick] (l4) -- (r2);
    \filldraw [white] (2.625,-2.5) circle (6pt);
    \draw[thick] (l1) -- (r3);
    \filldraw [white] (2,-2.2) circle (5pt);
    \filldraw [white] (2,-2.8) circle (5pt);
    \draw[thick] (l2) .. controls +(right:1cm) and +(right:1cm) ..(l3);
    \draw[thick] (r1) .. controls +(left:2cm) and +(left:2cm) ..(r4);
\end{tikzpicture}
\end{array}\\
\end{aligned}
\label{eq:definition of A_i}
\end{equation}
In these diagrams, the contours are arranged in the order $1,\one,2,\two$ from top to bottom (see diagram of $A_1$ in~(\ref{eq:definition of A_i})). We note that $\mathcal{L}_2=-2A_1+N^{-1}A_2-N^{-1}A_3 \equiv \sum_i v^{(1)}_i A_i$. We also define $\mathcal{L}_2^n=\sum_i v^{(n)}_i A_i$. We can work out the matrix representation of $\mathcal{L}_2$ on this basis. Define $\mathcal{L}_2 A_i=\sum_j M_{ji} A_j$, the $8 \times 8$ representation matrix $M$ is explicitly given by:
\begin{equation}
M=\begin{pmatrix}
-2 & 0 & -2N^{-1} & 0 & 0 & 0 & 0 & 0 \\
N^{-1} & -1 & 0 & 0 & 0 & 0 & 0 & 0 \\
-N^{-1} & 0 & -2 & 0 & 0 & -N^{-1} & 0 & 0 \\
0 & 0 & N^{-1} & -1 & 0 & 0 & 0 & 0 \\
0 & 2N^{-1} & 0 & 0 & 0 & 0 & 0 & 2N^{-1} \\
0 & 0 & -2N^{-1} & 0 & 0 & -2 & 0 & 0 \\
0 & 0 & 0 & 4N^{-1} & 0 & 0 & 0 & 0 \\
0 & 0 & 0 & 0 & 0 & N^{-1} & 0 & -1 \\
\end{pmatrix}
\label{eq:matrix repre in 2-replica}
\end{equation}
From the form of $M$, we immediately observe the following:
\begin{itemize}
\item[1.]\textit{Steady correlation patterns:} $\mathcal{L}_2 A_5 = \mathcal{L}_2 A_7 = 0$, meaning that these two bases are stationary. They directly correspond to the correlation patterns of the Haar ensemble~\cite{enwiki:1183923474,Collins_2022,Gu2013MomentsOR,zhang2015matrix,Czech:2023rbh}, which persist in the $t \rightarrow \infty$ limit.
\item[2.]\textit{Transient correlation patterns:} The off-diagonal entries of $M$ are suppressed by large $N$, so for large $N$, only the diagonal entries (contributed solely from $(P_{1\one}+P_{2\two}+P_{1\two}+P_{2\one})$, with $N^{-1}(X_{12}+X_{\one\two})$ only contributing to off-diagonals) are significant, which are of order one. Therefore, the diagonals are all negative for $A_i, i \in \{1,2,3,4,6,8\}$, indicating that these correlation patterns eventually fade away. This holds for general $k$. The correlation patterns connecting the left ($t=t_{\text{final}}$) and right boundary ($t=0$) will fade away, leaving only left/left or right/right patterns, which sum up to be the projectors of the Haar ensemble. This conjecture for general $k$ will be further clarified in section~\ref{section:low energy spectrum}.
\item[3.]\textit{Pseudo-Hermicity:} $\mathcal{L}_2$ is a Hermitian operator, but its representation matrix $M$ is non-Hermitian. This is because the basis $\{A_i\}$ is not orthogonal, making $M$ pseudo-Hermitian.
\end{itemize}

Knowing the $8 \times 8$ matrix $M$ and the $8 \times 1$ column vector $v^{(1)}$, we have:
\begin{equation}
e^{\mathcal{L}_2 t} = A_1 + \sum_{i=1}^{8}\left[(e^{Mt} - 1)M^{-1} v^{(1)}\right]_i A_i
\end{equation}
where $M^{-1}$ is interpreted as the pseudo-inverse of $M$, since $M$ itself is not invertible due to two zero modes.

It is instructive to study the spectrum of $M$. The eight eigenvalues $\lambda_i$ are given by:
\begin{equation}
\lambda_i = 0, 0, -1, -1, -1, -(2 - 2N^{-1}), -2, -(2 + 2N^{-1})
\end{equation}
We see that the $0, -1, -2$ eigenvalues are expected since the four-contour evolution is schematically the square of two-contour evolution, where the latter has $\{1, e^{-t}\}$ modes, so its square naturally includes $\{1, e^{-t}, e^{-2t}\}$ modes. The new modes are $\lambda_i = -2 \pm 2N^{-1}$, which have non-trivial $N$-dependence. This effect is non-perturbative.

Denoting $V$ as the eigenvector matrix of $M$ (each column of $V$ is a left-eigenvector of $M$, which does not need to be orthogonal) and $D$ as the diagonal matrix of eigenvalues of $M$, we can easily exponentiate $M$, and the final result is:
\begin{equation}
\mathcal{U}_2(t) = e^{\mathcal{L}_2 t} = \sum_{i=1}^{8} \left[\delta_{i1} + V(e^{Dt} - 1) V^{-1} M^{-1} v^{(1)}\right]_i A_i \equiv \sum_{i=1}^8 f_i(t) A_i
\label{eq:U_2}
\end{equation}

The time-dependent coefficients $f_i(t)$ in front of each correlation pattern are given explicitly by:
\begin{equation}
\begin{pmatrix}
f_1(t) \\
f_2(t) \\
f_3(t) \\
f_4(t) \\
f_5(t) \\
f_6(t) \\
f_7(t) \\
f_8(t) \\
\end{pmatrix}
=
\begin{pmatrix}
0 & 0 & \frac{1}{4} & \frac{1}{2} & \frac{1}{4}\\
0 & \frac{N^2-2}{N(N^2-4)} & \frac{-1}{4(N-2)} & \frac{-1}{2N} & \frac{-1}{4(N+2)}\\
0 & 0 & -\frac{1}{4} & 0 & \frac{1}{4}\\
0 & \frac{-1}{N^2-4} & \frac{1}{4(N-2)} & 0 & \frac{-1}{4(N+2)}\\
\frac{1}{N^2-1} & \frac{-2}{N^2-4} & \frac{1}{2(N-1)(N-2)} & 0 & \frac{1}{2(N+1)(N+2)}\\
0 & 0 & \frac{1}{4} & -\frac{1}{2} & \frac{1}{4}\\
\frac{-1}{N^3-N} & \frac{4}{N(N^2-4)} & \frac{-1}{2(N-1)(N-2)} & 0 & \frac{1}{2(N+1)(N+2)}\\
0 & \frac{2}{N(N^2-4)} & \frac{-1}{4(N-2)} & \frac{1}{2N} & \frac{-1}{4(N+2)}\\
\end{pmatrix}
\times
\begin{pmatrix}
1 \\
e^{-t}\\
e^{-(2-2N^{-1})t}\\
e^{-2t}\\
e^{-(2+2N^{-1})t}\\
\end{pmatrix}
\label{eq:define f_i(t)}
\end{equation}
where we express it in a compact way by decomposing it into five distinct modes.
\begin{figure}[t]
    \centering
    \includegraphics[width=0.50\textwidth]{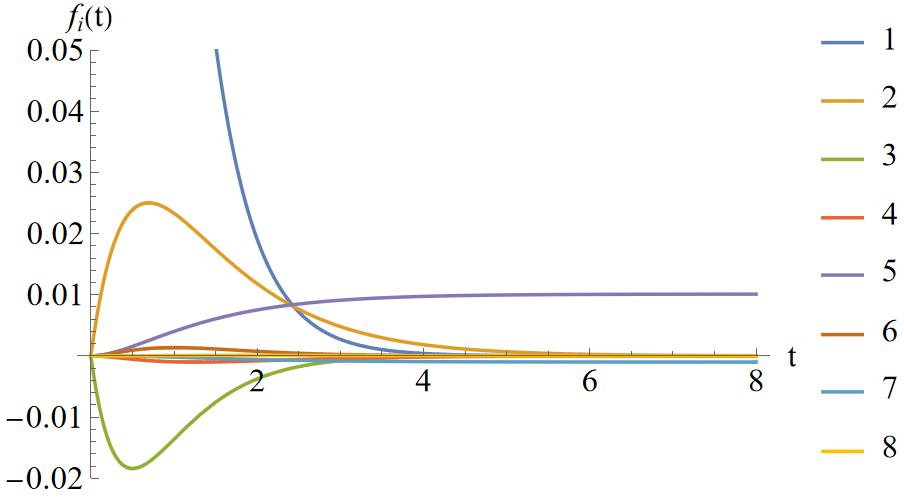}
    \includegraphics[width=0.49\textwidth]{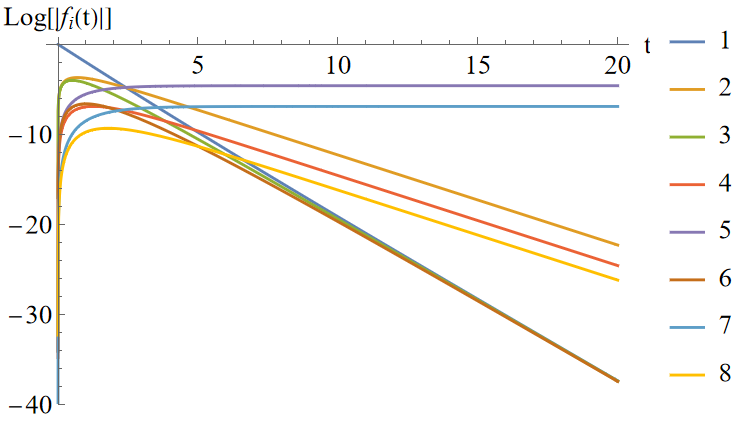}
    \caption{The pattern coefficient $f_i(t)$ or its log-plot, evaluated at $N=10$ according to equation~(\ref{eq:define f_i(t)}). From the log-plot, we can easily see the exponentially decaying modes, and $f_5(t),f_7(t)$ remain finite at $t=+\infty$.}
    \label{fig:pattern coefficient}
\end{figure}

We can perform some explicit sanity checks: (1) the first column of the coefficient matrix in~(\ref{eq:define f_i(t)}) shows that only $f_5,f_7$ are non-zero at $t=+\infty$. The two non-zero entries of first column, $\frac{1}{N^2-1}$ and $\frac{-1}{N^3-N}$, coincide with the correct Weingarten functions of the Haar ensemble; (2) one can check that the normalization is correct by noting that we should have $\tr[U_t U_t^{\dagger}] \tr[U_t U_t^{\dagger}] = N^2$, which is realized by $\langle \text{EPR}_{1\one} \text{EPR}_{2\two}|\mathcal{U}_4|\text{EPR}_{1\one}\text{EPR}_{2\two}\rangle = \sum_i f_i(t) \tr[A_i]$; (3) At $t=0$, only $f_1=1$ and other coefficients are zero.

In figure~\ref{fig:pattern coefficient}, we plot the correlation pattern coefficient $f_i(t)$ as a function of time. We see that some transient correlation patterns that connect the left-boundary condition and right-boundary condition may appear at intermediate times. However, they eventually die out at long real-time, leaving only $A_5$ and $A_7$:
\begin{equation}
\mathcal{U}_2(+\infty) = f_5(+\infty) A_5 + f_7(+\infty) A_7 = \int_{\text{Haar}} dU \cdot U \otimes U^* \otimes U \otimes U^*
\end{equation}
This means that $\mathcal{U}_2(+\infty)$ equals the prediction of the Haar ensemble~\cite{enwiki:1183923474,Collins_2022,Gu2013MomentsOR,zhang2015matrix,Czech:2023rbh}. In section~\ref{section:low energy spectrum}, we will show that $\mathcal{U}_k(t=+\infty)$ equals the Haar's prediction for arbitrary $k$.

We are now ready to apply the result of $\mathcal{U}_1(t)$ in (\ref{eq:U_1}) and the result of $\mathcal{U}_2(t)$ in (\ref{eq:U_2}), (\ref{eq:definition of A_i}), and (\ref{eq:define f_i(t)}) to calculate observables.

\subsection{One-replica observables}
\label{section:1 replica observable}
\paragraph{Spectral Form Factor.}
The spectral form factor (SFF) is given by:
\begin{equation}
\label{eq:SFF}
\E[\text{SFF}(t)] \equiv \E\left[\tr(U_t) \tr(U_t^{\dagger})\right] = N^2 e^{-t} + (1 - e^{-t})
\end{equation}
which smoothly interpolates between the identity and Haar ensemble. We see that at the timescale $t \sim O(\log N^2)$, the two terms become comparable, and SFF is of order one. 

\begin{figure}[t]
    \centering
    \includegraphics[width=0.51\textwidth]{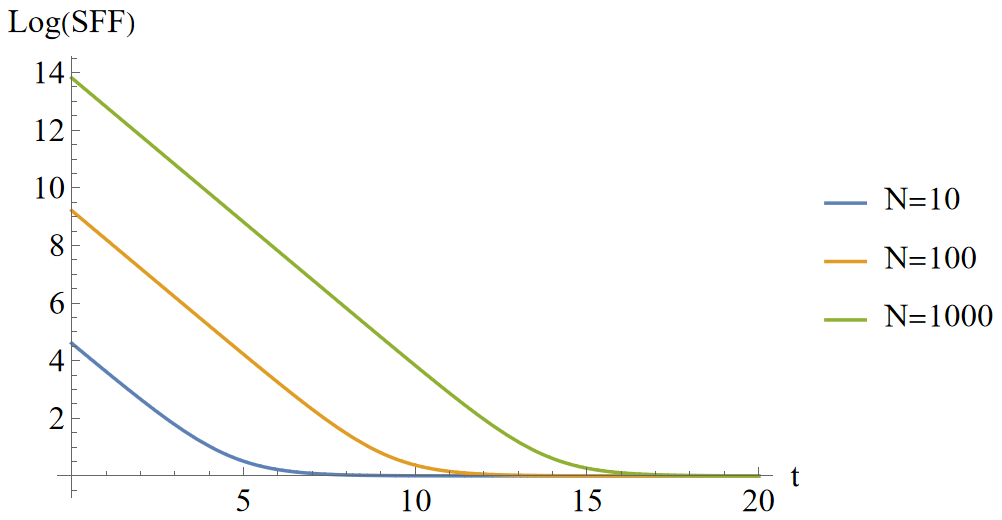}
    \includegraphics[width=0.47\textwidth]{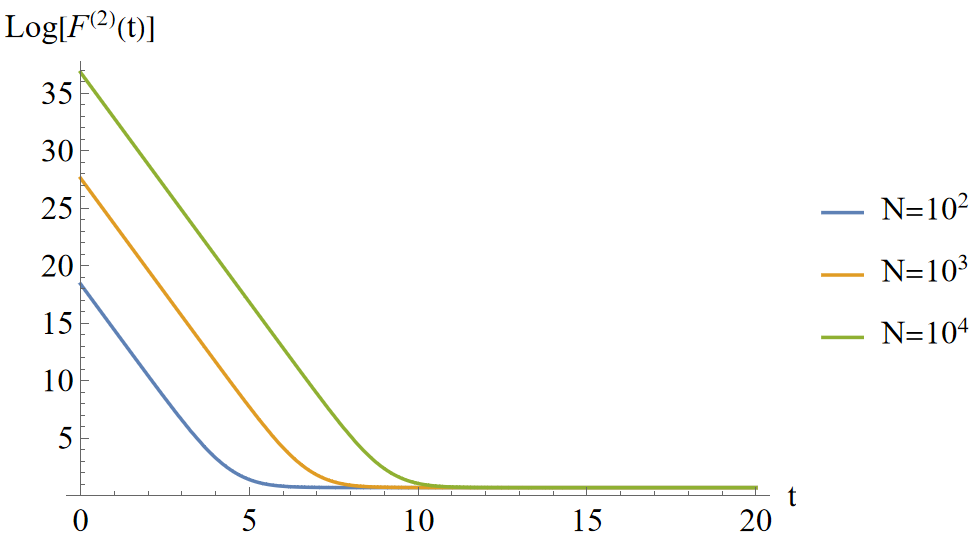}
    \caption{\textbf{Left:} Log of the average value of the spectral form factor~(\ref{eq:SFF}) with respect to time. \textbf{Right:} Log of the frame potential~(\ref{eq:frame potential}) with respect to time.}
    \label{fig:SFF and frame potential}
\end{figure}

Similar to the Brownian SYK, SFF in BGUE monotonically decreases with time, and there is no linear ramp as in the static GUE or static SYK model~\cite{Cotler2017BHandRMT}. In figure~\ref{fig:SFF and frame potential}, we plot SFF with respect to time.

\paragraph{Two-point function.}
Consider the following two-point function:
\begin{equation}
\label{eq:2pt}
\E[C(t)] \equiv \E[\tr(A B_t)] =\E[\tr(A U_t B U_t^{\dagger})] = \tr(AB) e^{-t} + N^{-1} \tr(A) \tr(B) (1 - e^{-t})
\end{equation}
which smoothly interpolates between $\tr(AB)$ and $\tr(A) \tr(B)$. This indicates that the correlation connecting the left and right of the contour gradually dies out at the timescale $t \sim O(1)$.

It is helpful to compare this with the result of BDSSYK (Brownian Double-Scaled-SYK), whose two-point function has the form of $e^{-\frac{J^2}{\lambda}(1-q_V)t}$~\cite{Milekhin:2023bjv,Stanford:2023npy}. There are several immediate remarks:
\begin{itemize}
\item[1.]\textit{Emergence of tomperature in BGUE.} The emergence of a timescale in infinite temperature (Brownian has no energy conservation) is like tomperature~\cite{SusskindLin2022nss}.
\item[2.]\textit{Replica-wormhole in BGUE.} There is no replica wormhole in BDSSYK. This is because we already take the infinite dimension limit, and all higher topological changes are suppressed. In our BGUE model, the non-decaying part of the two-point function~\cite{Stanford:2020wkf,Saad:2019pqd,JuanMaldacena_2003} $\E[C(t=+\infty)] = N^{-1} \tr(A) \tr(B)$ is reminiscent of the replica wormhole~\cite{Penington2022,Almheiri2020}.
\item[3.]\textit{Dependence on matter scaling dimension.} In BDSSYK, the scaling dimension of matter chords, $q_V$, appears \textit{in} the exponent, which affects the timescale explicitly. In BGUE, the matter operators only appear as the boundary condition, and the damping mode only depends on $\mathcal{L}_2$. BGUE is recovered as a special limit in BDSSYK by taking the $q$-parameter to zero.
\end{itemize}

\subsection{Two-replica observables}
\label{section:2 replica observable}
\paragraph{Fluctuation of spectral form factor.}
\begin{equation}
\label{eq:variance of SFF}
\begin{aligned}
\text{Var}[\text{SFF}(t)] &\equiv \E[\tr(U_t) \tr(U_t^{\dagger}) \tr(U_t) \tr(U_t^{\dagger})] - \E[\tr(U_t) \tr(U_t^{\dagger})] \E[\tr(U_t) \tr(U_t^{\dagger})] \\
&= 1 + 2(N^2 - 1) e^{-t} + N^2 e^{-2t} \left(N^2 \sinh^2{\frac{t}{N}} - N \sinh{\frac{t}{N}} - \frac{3}{2} \cosh{\frac{2t}{N}} - \frac{1}{2} + N^{-2}\right)
\end{aligned}
\end{equation}
We notice that $\text{Var}[\text{SFF}(0)] = 0$, which is reasonable, and $\text{Var}[\text{SFF}(+\infty)] = 1$. We can also fix $t$ and scale $N$ to infinity:
\begin{equation}
\lim_{N \rightarrow \infty} \text{Var}[\text{SFF}(t)] = N^2 \left[2 e^{-t} + e^{-2t}(t^2 - t - 2)\right]
\end{equation}
Therefore, the maximal value of variance is of order $O(N^2)$, and the timescale for reaching this maximal fluctuation is of order one.

\begin{figure}[t]
    \centering
    \includegraphics[width=0.43\textwidth]{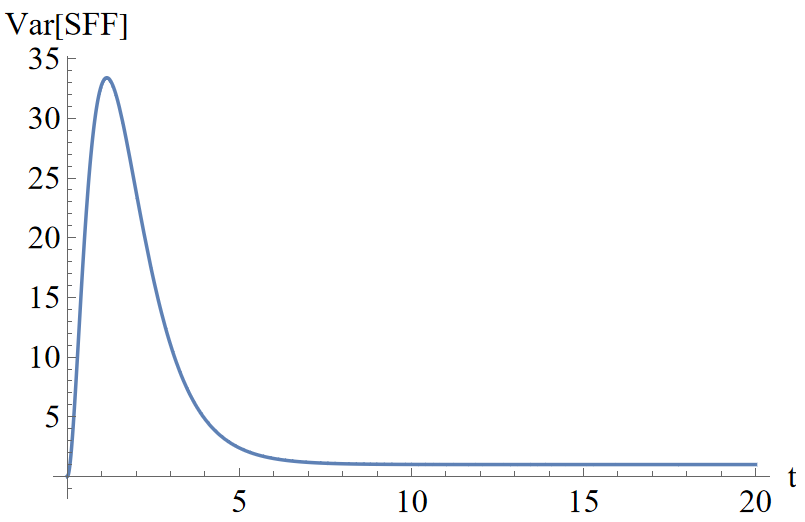}
    \includegraphics[width=0.56\textwidth]{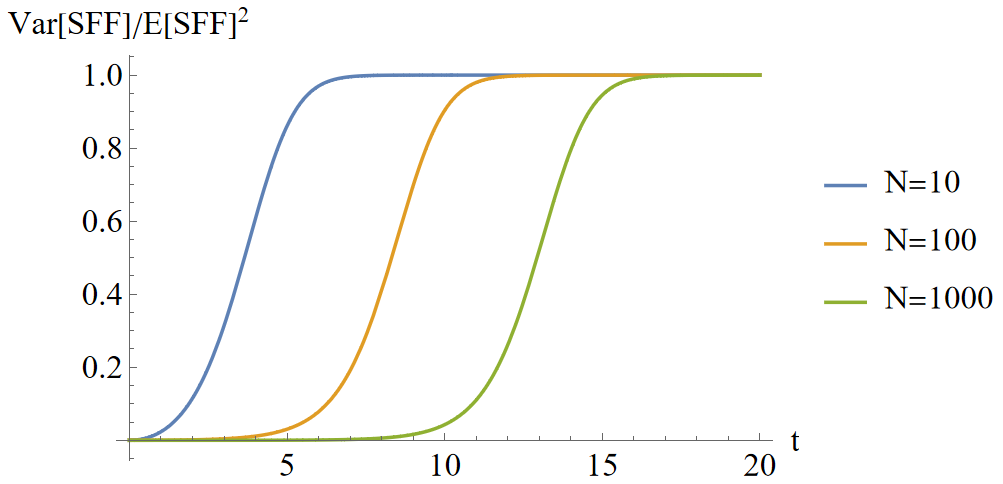}
    \caption{\textbf{Left:} The absolute variance of the spectral form factor~(\ref{eq:variance of SFF}), dimension chosen at $N=10$. \textbf{Right:} The relative variance of SFF.}
    \label{fig:variance of SFF}
\end{figure}

A more physical way of quantifying the size of the fluctuation is to calculate the relative variance: $\eta(t) \equiv \text{Var}[\text{SFF}(t)] / \E[\text{SFF}(t)]^2$, which is shown in figure~\ref{fig:variance of SFF}. We see that for $t < O(\log N^2)$, $\eta \approx 0$, and for $t > O(\log N^2)$, $\eta \approx 1$. This is consistent with the case in GUE or SYK, where the fluctuation in the plateau regime is large compared to the dip regime~\cite{Cotler2017BHandRMT}.

\paragraph{Fluctuation of two-point functions.}
In general, the expression for the fluctuation of the two-point function is complicated, as calculating $\E[\tr(AB_t) \tr(AB_t)]$ involves many contour pairings, where each of the $A_i$ terms will contribute respectively as: $A_1 \rightarrow \tr[AB]^2$, $A_2 \rightarrow 2 \tr[AB] \tr[A] \tr[B] + 2 \tr[A^2 B^2]$, $A_3 \rightarrow 2 \tr[ABAB]$, $A_4 \rightarrow 4 \tr[B] \tr[BA^2] + 4 \tr[A] \tr[AB^2]$, $A_5 \rightarrow \tr[A]^2 \tr[B]^2 + \tr[A^2] \tr[B^2]$, $A_6 \rightarrow \tr[AB]^2$, $A_7 \rightarrow \tr[A]^2 \tr[B^2] + \tr[B]^2 \tr[A^2]$, $A_8 \rightarrow 2 \tr[A] \tr[B] \tr[AB] + 2 \tr[A^2 B^2]$. To simplify the expression, we consider three cases:
\begin{itemize}
\item[1.]\textit{Case 1:} $\tr[A] = \tr[B] = \tr[AB] = 0, A^2 = B^2 = 1, AB = BA$, where we imagine $A, B$ are Pauli operators on different spatial sites. In this case, we have $\E[C_1(t)] = 0$, meaning the average two-point function vanishes.
\item[2.]\textit{Case 2:} $\tr[A] = \tr[B] = \tr[AB] = 0, A^2 = B^2 = 1, AB = -BA$, where we imagine $A, B$ are Pauli operators on the same spatial site, but with different flavors. In this case, we have $\E[C_2(t)] = 0$, meaning the average two-point function vanishes.
\item[3.]\textit{Case 3:} $A = B, \tr[A] = 0, A^2 = 1$, where we consider the auto-correlation of a Pauli operator. In this case, we have $\E[C_3(t)] = N e^{-t}$.
\end{itemize}
\begin{figure}[t]
    \centering
    \includegraphics[width=0.55\textwidth]{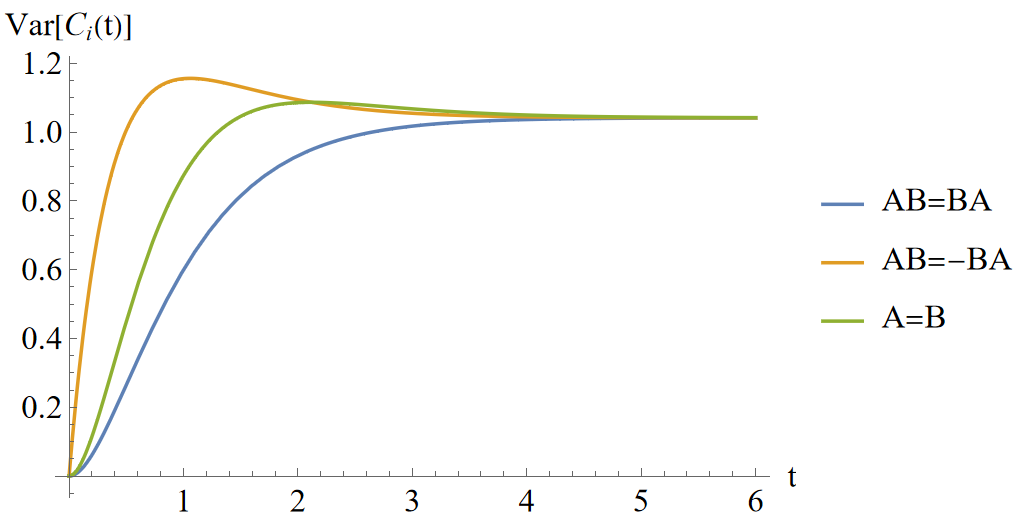}
    \includegraphics[width=0.44\textwidth]{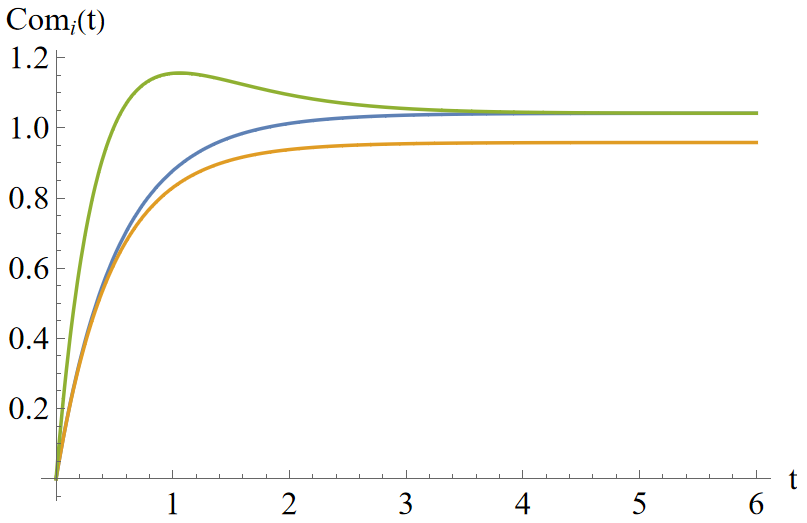}
    \caption{\textbf{Left:} Fluctuation of two-point functions~(\ref{eq:variance of 2pt}). \textbf{Right:} Average value of commutator squared~(\ref{eq:OTOC}). In both figures, the dimension is chosen at $N=5$.}
    \label{fig:fluctuation of 2pt and commutator}
\end{figure}
The variance of these three types of correlation functions is given by:
\begin{equation}
\label{eq:variance of 2pt}
\begin{aligned}
\text{Var}[C_1(t)] &= \frac{N^2}{N^2-1} - \frac{N^2}{2(N-1)} e^{-(2-2N^{-1})t} + \frac{N^2}{2(N+1)} e^{-(2+2N^{-1})t}\\
\text{Var}[C_2(t)] &= \frac{N^2}{N^2-1} + \frac{N(N-2)}{2(N-1)} e^{-(2-2N^{-1})t} - \frac{N^2}{N(N+2)} e^{-(2+2N^{-1})t}\\
\text{Var}[C_3(t)] &= \frac{N^2}{N^2-1} + \frac{N^2(N-2)}{2(N-1)} e^{-(2-2N^{-1})t} - N^2 e^{-2t} + \frac{N^2(N+2)}{2(N+1)} e^{-(2+2N^{-1})t}\\
\end{aligned}
\end{equation}
We notice that the variance is of order one at arbitrary $t, N$. This finite variance is non-trivial for $C_1(t)$ and $C_2(t)$, since their average values vanish. This non-zero variance with zero mean is due to the effect of the replica wormhole, as indicated in~\cite{Stanford:2020wkf}.

These three fluctuations are shown in figure~\ref{fig:fluctuation of 2pt and commutator}, where we see that at intermediate times, the behavior of fluctuation depends on whether $A, B$ commute or anti-commute, and also on whether $A, B$ are the same or not. This originates from terms like $\tr[ABAB], \tr[AB]$, whose coefficients are non-zero only at intermediate times. 

We also notice that the initial growth of variance differs in the three cases. To see this, we may work out the Taylor expansion at short times:
\begin{equation}
\begin{aligned}
\text{Var}[C_1(t)] &= 2t^2 - \frac{8}{3} t^3 + \frac{2(3N^2 + 1)}{3N^2} t^4 + O(t^5)\\
\text{Var}[C_2(t)] &= 4t - 6t^2 + \frac{8(2N^2 + 1)}{3N^2} t^3 + O(t^4)\\
\text{Var}[C_3(t)] &= 4t^2 - \frac{20}{3} t^3 + \frac{2(9N^2 + 1)}{3N^2} t^4 + O(t^5)\\
\end{aligned}
\end{equation}
It is interesting to see that for $C_2$, the variance grows linearly while the others grow quadratically. It is also interesting to observe that the first two coefficients are independent of $N$.

We can also work out the limit where $N$ goes to infinity while keeping $t$ fixed, where the variance is of order one:
\begin{equation}
\begin{aligned}
\lim_{N \rightarrow \infty} \text{Var}[C_1(t)] &= 1 + e^{-2t}(-2t - 1)\\
\lim_{N \rightarrow \infty} \text{Var}[C_2(t)] &= 1 + e^{-2t}(2t - 1)\\
\lim_{N \rightarrow \infty} \text{Var}[C_3(t)] &= 1 + e^{-2t}(2t^2 - 2t - 1)\\
\end{aligned}
\end{equation}

\paragraph{Out-of-Time-Ordered Correlator (OTOC)}
Assuming $A^2 = B^2 = 1$, the relation between (anti)commutator-squared and OTOC is given by:
\begin{equation}
\text{Com}(t) \equiv (2N)^{-1} \tr\left([A, B_t]^{\dagger}_{\pm} [A, B_t]_{\pm}\right) = 1 \pm N^{-1} \tr\left(AB_t AB_t\right) = 1 \pm \text{OTOC}(t)
\end{equation}
where we choose (anti)commutator and (plus)minus sign on RHS if at the initial time $A, B$ (anti)commute with each other. The commutator of three cases is given by:
\begin{equation}
\label{eq:OTOC}
\begin{aligned}
\E[\text{Com}_1(t)] &= 1 + \frac{1}{N^2 - 1} - \frac{N}{2(N-1)} e^{-(2-2N^{-1})t} - \frac{N}{2(N+1)} e^{-(2+2N^{-1})t}\\
\E[\text{Com}_2(t)] &= 1 - \frac{1}{N^2 - 1} - \frac{N-2}{2(N-1)} e^{-(2-2N^{-1})t} - \frac{N+2}{2(N+1)} e^{-(2+2N^{-1})t}\\
\E[\text{Com}_3(t)] &= 1 + \frac{1}{N^2 - 1} + \frac{N(N-2)}{2(N-1)} e^{-(2-2N^{-1})t} - \frac{N(N+2)}{2(N+1)} e^{-(2+2N^{-1})t}\\
\end{aligned}
\end{equation}
We may take $N \rightarrow \infty$ while keeping $t$ finite:
\begin{equation}
\begin{aligned}
\lim_{N \rightarrow \infty} \E[\text{Com}_1(t)] &= 1 - e^{-2t}\\
\lim_{N \rightarrow \infty} \E[\text{Com}_2(t)] &= 1 - e^{-2t}\\
\lim_{N \rightarrow \infty} \E[\text{Com}_3(t)] &= 1 + e^{-2t}(2t - 1)\\
\end{aligned}
\end{equation}
We note that OTOC and commutators still have non-trivial dynamics in the infinite $N$ limit. This hyper-fast scrambling behavior~\cite{Susskind2021omt,Susskind2021dfc,Susskind2021esx,Susskind2022fop,Susskind2022dfz,SusskindLin2022nss,Susskind2022bia,Susskind2023hnj,Susskind2023rxm} is consistent with other Brownian models such as BDSSYK~\cite{Milekhin:2023bjv,Stanford:2023npy}. This is opposed to static SYK, where the timescale for scrambling is $O(\log N)$~\cite{Maldecena2016}, meaning no scrambling dynamics will occur at infinite $N$. The behavior of commutators is shown in figure~\ref{fig:fluctuation of 2pt and commutator}.

\begin{figure}
    \centering
    \includegraphics[width=0.51\textwidth]{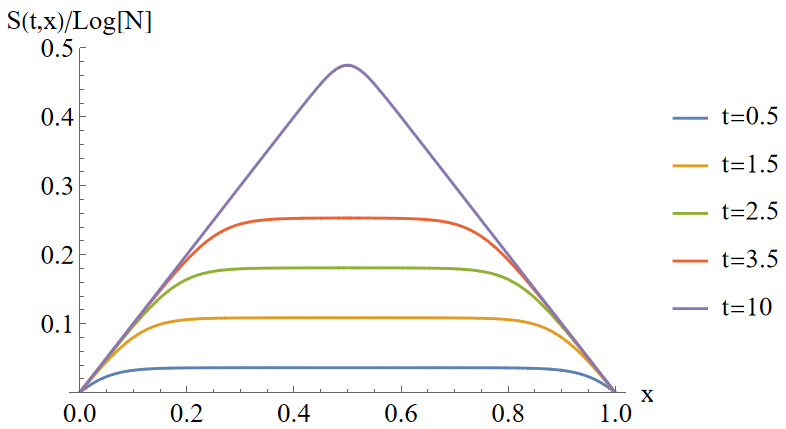}
    \includegraphics[width=0.48\textwidth]{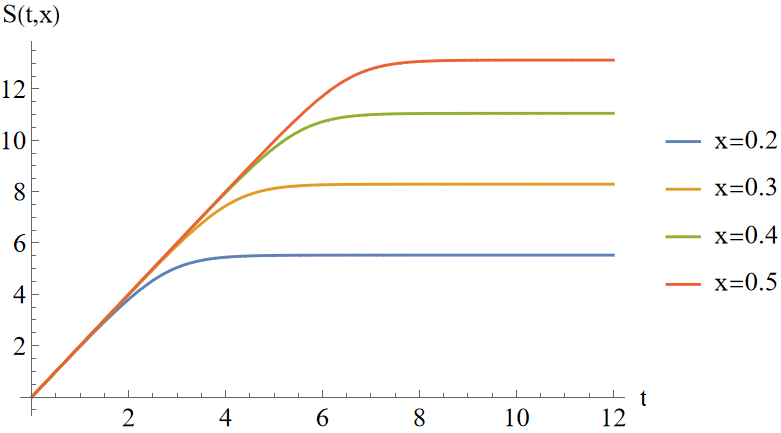}
    \caption{The entanglement entropy~(\ref{eq:renyi entropy}) changes with respect to subsystem size and time. Size is taken to be $N=10^{12}$.}
    \label{fig:entropy}
\end{figure}

\paragraph{Second Rényi Entropy}
Consider we evolve the system from a pure product state. At time $t$, we consider the sub-system with Hilbert space dimension $N^x, x \in [0, 1]$. The second Rényi entropy (actually we first average over the purity, and then take the log) is given by:
\begin{equation}
\label{eq:renyi entropy}
S^{(2)}(t, x) = -\log\left[e^{-(2 + 2N^{-1})t} + \frac{N^x + N^{1-x}}{N+1}\left(1 - e^{-(2 + 2N^{-1})t}\right)\right]
\end{equation}
We see that the system approaches the Page curve of the Haar random ensemble at order one time, consistent with studies in the scrambling phase of MIPT in (0+1)-dimensional models~\cite{koh2022experimental,li2022cross,MIPT1,MIPT10,MIPT2,MIPT24,MIPT3,MIPT30,MIPT5,MIPT7,noel2022measurement}.

We can also study the velocity of entropy growth at the initial time:
\begin{equation}
S^{(2)}(t, x) = 2(1 - N^{-x})(1 - N^{-(1-x)})t - 2(1 - N^{-x})(1 - N^{-(1-x)})(N^{-x} + N^{-(1-x)})t^2 + O(t^3)
\end{equation}
We see that the coefficient of $t$ approaches $2$ for arbitrary $x$ in the infinite $N$ limit, and the coefficient of $t^2$ vanishes in the infinite $N$ limit. This shows that in the infinite $N$ limit, the linear-in-time growth rate of entropy for all subsystems is the same: 
\begin{equation}
\lim_{N \rightarrow \infty} S^{(2)}(t, x) = 2t, \ t < O(\log N)
\end{equation}

Such a growth rate can be interpreted from (\ref{eq:renyi entropy}). We notice that for $x \neq 0, x \neq 1$ and small $t$ before saturation, the second term $\frac{N^x + N^{1-x}}{1 + N}$ is of order $O(N^{-\min(x, 1-x)}) \ll O(1)$, which is always subleading in the infinite $N$ limit compared to the first term. Therefore, we can safely approximate $S^{(2)}(t, x) \approx -\log[e^{-(2 + 2N^{-1})t}] = 2(1 + N^{-1})t \approx 2t$.
The entropy curve is shown in figure~\ref{fig:entropy}. 

\paragraph{Frame potential for unitary two-design}
Given any ensemble $\mE$ of unitary matrices, the frame potential~\cite{Scott_2008,Gross_2007,hunterjones2019unitary,PRXQuantum.2.030316,Jian:2022pvj,Cotler2017ChaosComplexityRMT,Roberts2017ChaosByDesign}, relevant to $k$-designs, is defined as:
\begin{equation}
F^{(k)}_{\mE}(t) = \E_{U, V \in \mE}\left[ \left|\tr(U V^{\dagger})\right|^{2k}\right]
\end{equation}
where $U$ and $V$ are drawn independently from the same ensemble $\mE$. 

Frame potential naturally measures whether $\mathcal{E}$ is 'dispersed' or 'concentrated'. Consider an extreme case where the ensemble consists of only one element with probability one. In this case, the forward evolution perfectly cancels the backward evolution, leading to $\tr(UV^{\dagger}) = \tr(\mathbb{I}) = N$, which results in the maximal possible value of the frame potential. On the other hand, if $\mathcal{E}$ is dispersed, then two independently drawn samples $U$ and $V$ are likely to be very different, meaning that $\tr(UV^\dagger)$ will be very small. The most 'dispersed' ensemble is the Haar ensemble, indicating that $F^{(k)}_{\text{Haar}}$ is the smallest among all $\mathcal{E}$.

In our case, we can calculate $k=2$, which is relevant for comparing 2-designs. For the identity ensemble (an ensemble whose single element is the identity matrix), $F^{(2)}_{\text{id}} = N^4$ and for the Haar random ensemble we have $F^{(2)}_\Haar = 2! = 2$. 

The formula for the frame potential of BGUE is $F_\BGUE^{(2)}(t) = \tr[e^{\mathcal{L}_2 t} \times e^{\mathcal{L}_2 t}] = \tr[e^{\mathcal{L}_2 2t}] = \E[\text{SFF}(2t)\text{SFF}(2t)]$, which is the square of the average spectral form factor. The explicit formula is given by:
\begin{equation}
\label{eq:frame potential}
\begin{aligned}
F^{(2)}_\BGUE(t) = 2 + 4(N^2 - 1)e^{-2t} &+ \frac{1}{4} N^2 (N + 1)(N - 3) e^{-(4 - 4N^{-1})t} + \frac{1}{2} (N^2 - 1)(N^2 - 4) e^{-4t} \\
&+ \frac{1}{4} N^2 (N - 1)(N + 3) e^{-(4 + 4N^{-1})t}
\end{aligned}
\end{equation}
which smoothly interpolate between $F^{(2)}_{\text{id}}$ and $F^{(2)}_\Haar$. In section~\ref{section:calculating frame potential}, we obtain the frame potential for arbitrary $k$ up to the second excited states of $\mathcal{L}_k$. For $k=2$, the second excited states are already the highest energy states.

\section{Approaching $k$-design for arbitrarily many replicas}
\label{section:approaching k-design}
In this section, we are interested in how $\mathcal{U}_k(t)$ approaches the Haar ensemble's prediction for large but finite $t$. Therefore, in section~\ref{section:low energy spectrum}, we focus on the low-energy spectrum (up to the family of second excited states) of $-\mathcal{L}_k$, which governs the long-time behavior of $\mathcal{U}_k(t)$. Using this spectral information, in section~\ref{section:calculating frame potential}, we calculate the frame potential $F^{(k)}(t)$ and estimate a lower bound on the time needed to approach an $\varepsilon$-approximated $k$-design.

Recall there are $k$-replicas with $k$ forward-moving contours labeled by $i=1,2,\cdots,k$ and $k$ backward-moving contours labeled by $\bar{i}=1,2,\cdots,k$. The Hilbert space is $(\mathbb{C}^{\otimes N})^{\otimes 2k}$. The averaged evolution operator is defined by:
\begin{equation}
\mathcal{U}_k(t)=\E\left[U_t^{\otimes k}\otimes U_t^{*\otimes k}\right]=e^{\mathcal{L}_kt}
\end{equation}
The Lindbladian is given by:
\begin{equation}
\begin{aligned}
\mathcal{L}_k&=-k+N^{-1}\sum_{i,\bar{j}=1}^kP_{i\bar{j}}-N^{-1}\sum_{1\leq i<j\leq k}X_{ij}-N^{-1}\sum_{1\leq \bar{i}<\bar{j}\leq k}X_{\bar{i}\bar{j}}\\
&\equiv-k+N^{-1}\mathcal{P}-N^{-1}\mathcal{X}-N^{-1}\bar{\mathcal{X}}
\end{aligned}
\end{equation}

\subsection{Low energy spectrum}
\label{section:low energy spectrum}

\subsubsection{Ground states}
In this section, we show that the ground state energy of $\mathcal{L}_k$ is zero, with degeneracy $k!$. The ground state manifold is spanned by permutation matrix elements. The projector onto the ground state manifold equals the prediction of the Haar average, indicating that $\mathcal{U}_k(t)$ approaches the Haar ensemble as $t \rightarrow +\infty$.

To be precise, the ground state of $\mathcal{L}_k$ is labeled by a permutation element $\pi\in\mathbb{S}^k$:
\begin{equation}
|\pi\rangle:\langle n_1... n_{i}... n_{k},n_{\bar{1}}...n_{\bar{i}}... n_{\bar{k}}|\pi\rangle=\delta_{n_1n_{\overline{\pi(1)}}}\cdots\delta_{n_in_{\overline{\pi(i)}}}\cdots\delta_{n_kn_{\overline{\pi(k)}}}
\end{equation}
where $n_i,n_{\bar{i}}=1,2,\cdots,N$ is the basis label on the individual Hilbert space of each contour. A diagrammatic example for $k=4$ is:
\begin{equation}
|\pi\rangle=
\begin{array}{c}
\begin{tikzpicture}[scale=0.35]
    \foreach \y in {1,2,3,4} {
        \coordinate (l\y) at (0, -\y);
    }
    \foreach \y in {5,6,7,8} {
        \coordinate (l\y) at (0, -\y-1);
    }

    \draw[thick,gray] (l1) .. controls +(right:3cm) and +(right:3cm) .. (l6);
    \draw[thick,gray] (l2) .. controls +(right:1cm) and +(right:1cm) .. (l5);
    \draw[thick,gray] (l3) .. controls +(right:2cm) and +(right:2cm) .. (l7);
    \draw[thick,gray] (l4) .. controls +(right:1cm) and +(right:1cm) .. (l8);

    \draw[lightgray] (l1)--(l4);
    \draw[lightgray] (l5)--(l8);

    \filldraw [lightgray] (0,-1) circle (4pt);
    \filldraw [lightgray] (0,-2) circle (4pt);
    \filldraw [lightgray] (0,-3) circle (4pt);
    \filldraw [lightgray] (0,-4) circle (4pt);
    \filldraw [lightgray] (0,-6) circle (4pt);
    \filldraw [lightgray] (0,-7) circle (4pt);
    \filldraw [lightgray] (0,-8) circle (4pt);
    \filldraw [lightgray] (0,-9) circle (4pt);
    
    \node[] at (0,-1) {};

    \node[text=gray] at (-0.7,-1) {\scriptsize 1};
    \node[text=gray] at (-0.7,-2) {\scriptsize 2};
    \node[text=gray] at (-0.7,-3) {\scriptsize 3};
    \node[text=gray] at (-0.7,-4) {\scriptsize 4};
    \node[text=gray] at (-0.7,-6) {\scriptsize $\bar{1}$};
    \node[text=gray] at (-0.7,-7) {\scriptsize $\bar{2}$};
    \node[text=gray] at (-0.7,-8) {\scriptsize $\bar{3}$};
    \node[text=gray] at (-0.7,-9) {\scriptsize $\bar{4}$};
\end{tikzpicture}
\end{array}
,\ \pi\in\mathbb{S}^4
\label{eq:pi}
\end{equation}

In these diagrams, the $2k$-contours are arranged from up to down in the order of $1,2,...,k,\one,\two,...,\bar{k}$ (see the label in~(\ref{eq:pi}) for example). Notice that in section~\ref{section:1/2 replica}, the diagram for equation~(\ref{eq:definition of A_i}) for $k=2$, the contour is ordered as $1,\one,2,\two$ from up to down.

Now, we are ready to show that $|\pi\rangle$ is indeed the eigenstate of $\mathcal{L}_k$ with zero energy. Let's act $\mathcal{L}_k$ on $|\pi\rangle$. One kind of term coming from $\mathcal{P}$ is:
\begin{equation}
\mathcal{P}|\pi\rangle\supset\adjincludegraphics[valign=c, width=0.07\textwidth]{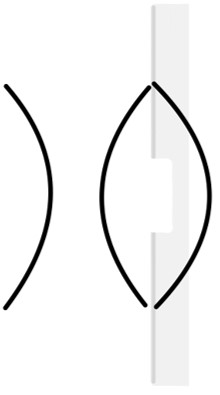}=N\cdot\adjincludegraphics[valign=c, width=0.024\textwidth]{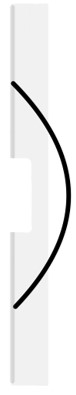}=N|\pi\rangle,\ \#=k
\end{equation}
Here, to simplify the diagram like~(\ref{eq:pi}), we use black lines to indicate the links that are relevant for calculation, and a gray background to represent the other links that are not involved in this step of the calculation.

\begin{table}[t]
    \centering
    \begin{tabular}{c|c|c}
    \hline Label & Equation & \# \\
    \hline \boxed{1} & $\adjincludegraphics[valign=c, width=0.1\textwidth]{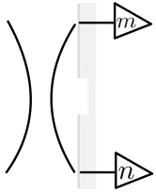}=\delta_{mn}\adjincludegraphics[valign=c, width=0.052\textwidth]{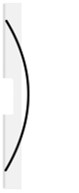}$ &  $1$ \\
    \hline \boxed{2} & $\adjincludegraphics[valign=c, width=0.1\textwidth]{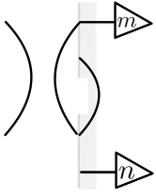}=\adjincludegraphics[valign=c, width=0.052\textwidth]{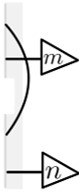}$ &  $k-1$ \\
    \hline \boxed{3} & $\adjincludegraphics[valign=c, width=0.1\textwidth]{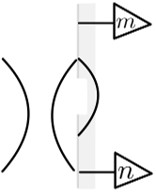}=\adjincludegraphics[valign=c, width=0.052\textwidth]{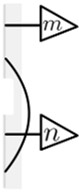}$ &  $k-1$ \\
    \hline \boxed{4} & $\adjincludegraphics[valign=c, width=0.1\textwidth]{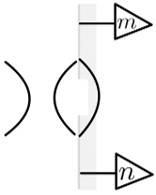}=N\cdot\adjincludegraphics[valign=c, width=0.052\textwidth]{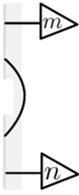}$ &  $k-1$ \\
    \hline \boxed{5} & $\adjincludegraphics[valign=c, width=0.1\textwidth]{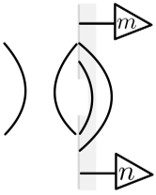}=\adjincludegraphics[valign=c, width=0.052\textwidth]{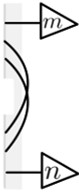}$ &  $2\binom{k-1}{2}$ \\
    \hline
    \end{tabular}
    \caption{ $\mathcal{P}|mn,i\bar{j},\pi\rangle$ is grouped into five categories. Notice that the total number of terms is $1+3(k-1)+2\binom{k-1}{2}=k^2$ is correct.}
    \label{table:first excited state, acting P}
\end{table}

The number of such terms is $\#=k$. So, this contribution cancels the $-k$ constant part in $\mathcal{L}_k$. The other terms in $\mathcal{P}$ look like:
\begin{equation}
\mathcal{P}|\pi\rangle\supset\adjincludegraphics[valign=c, width=0.07\textwidth]{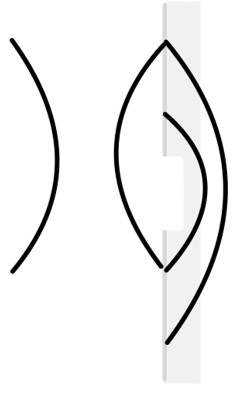}=\adjincludegraphics[valign=c, width=0.02\textwidth]{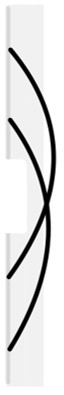},\ \mathcal{P}|\pi\rangle\supset\adjincludegraphics[valign=c, width=0.07\textwidth]{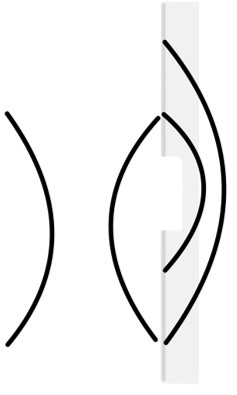}=\adjincludegraphics[valign=c, width=0.02\textwidth]{gs_5.jpg},\ \#=k^2-k
\end{equation}
They are canceled by the action of $\mathcal{X}$ and $\bar{\mathcal{X}}$:
\begin{equation}
\mathcal{X}|\pi\rangle\supset\adjincludegraphics[valign=c, width=0.07\textwidth]{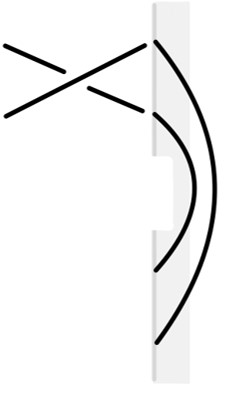}=\adjincludegraphics[valign=c, width=0.02\textwidth]{gs_5.jpg},\ \bar{\mathcal{X}}|\pi\rangle\supset\adjincludegraphics[valign=c, width=0.07\textwidth]{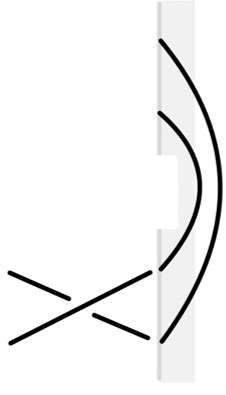}=\adjincludegraphics[valign=c, width=0.02\textwidth]{gs_5.jpg},\ \#=\binom{k}{2}+\binom{k}{2}=k^2-k
\end{equation}
We notice that the number of terms matches. Putting everything together, this shows that $\mathcal{L}_k|\pi\rangle=0$.

We notice that the set of ground states $\{|\pi\rangle|\pi\in \mathbb{S}^{k}\}$ are not orthogonal. In fact, if we define their Gram matrix as $g_{\sigma\tau}=\langle\sigma|\tau\rangle=N^{\ell(\sigma\tau^{-1})}$ ($\ell(\sigma\tau^{-1})$ counts the number of cycles in permutation element $\sigma\tau^{-1}$) and denote its (pseudo)inverse matrix as $g^{\sigma\tau}$ ($g^{\sigma\tau}$ is exactly the Weingarten function $\text{Wg}(\sigma,\tau)$~\cite{enwiki:1183923474,Collins_2022,Gu2013MomentsOR,zhang2015matrix,Czech:2023rbh}), then the ground state manifold projector $\mathcal{P}_{\text{GS}}^{(k)}$ is given by:
\begin{equation}
\mU_k(+\infty)=\mathcal{P}^{(k)}_{\text{GS}}=\sum_{\sigma,\tau\in\mathbb{S}^k}g^{\sigma\tau}|\sigma\rangle\langle\tau|=\int_{\text{Haar}}dU \cdot U^{\otimes k}\otimes U^{*\otimes k}
\end{equation}
This shows that BGUE evolves and saturates to the Haar ensemble at $t=+\infty$.

\subsubsection{First excited states}
In this section, we explicitly construct the wavefunction of the first excited states and show that their energy is $-1$ with degeneracy $k \cdot k! \cdot (N^2-1)$.

The first excited states are built upon some modifications of the ground state, where one of the links is broken, and two computational basis indices are glued onto the two broken ends. An example for $k=4$ is:

\begin{equation}
\begin{array}{c}
\begin{tikzpicture}[scale=0.35]
    \foreach \y in {1,2,3,4} {
        \coordinate (l\y) at (0, -\y);
    }
    \foreach \y in {5,6,7,8} {
        \coordinate (l\y) at (0, -\y-1);
    }

    \draw[thick,gray] (l1) .. controls +(right:3cm) and +(right:3cm) .. (l6);
    \draw[thick,gray] (l2) .. controls +(right:1cm) and +(right:1cm) .. (l5);
    \draw[thick,gray] (l3) .. controls +(right:2cm) and +(right:2cm) .. (l7);
    \draw[thick,gray] (l4) .. controls +(right:1cm) and +(right:1cm) .. (l8);

    \draw[lightgray] (l1)--(l4);
    \draw[lightgray] (l5)--(l8);

    \filldraw [lightgray] (0,-1) circle (4pt);
    \filldraw [lightgray] (0,-2) circle (4pt);
    \filldraw [lightgray] (0,-3) circle (4pt);
    \filldraw [lightgray] (0,-4) circle (4pt);
    \filldraw [lightgray] (0,-6) circle (4pt);
    \filldraw [lightgray] (0,-7) circle (4pt);
    \filldraw [lightgray] (0,-8) circle (4pt);
    \filldraw [lightgray] (0,-9) circle (4pt);

\end{tikzpicture}
\end{array}
\longrightarrow
\begin{array}{c}
\begin{tikzpicture}[scale=0.35]
    \foreach \y in {1,2,3,4} {
        \coordinate (l\y) at (0, -\y);
    }
    \foreach \y in {5,6,7,8} {
        \coordinate (l\y) at (0, -\y-1);
    }

    \draw[thick,gray] (l1) .. controls +(right:3cm) and +(right:3cm) .. (l6);
    \draw[thick,gray] (l2) .. controls +(right:1cm) and +(right:1cm) .. (l5);
    \draw[thick,gray] (l4) .. controls +(right:1cm) and +(right:1cm) .. (l8);

    \draw[lightgray] (l1)--(l4);
    \draw[lightgray] (l5)--(l8);

    \filldraw [lightgray] (0,-1) circle (4pt);
    \filldraw [lightgray] (0,-2) circle (4pt);
    \filldraw [lightgray] (0,-3) circle (4pt);
    \filldraw [lightgray] (0,-4) circle (4pt);
    \filldraw [lightgray] (0,-6) circle (4pt);
    \filldraw [lightgray] (0,-7) circle (4pt);
    \filldraw [lightgray] (0,-8) circle (4pt);
    \filldraw [lightgray] (0,-9) circle (4pt);

    \draw[thick] (l3) -- (3,-3);
    \draw[thick] (l7) -- (3,-8);

    \draw[thick] (3,-2.2)--(3,-3.8);
    \draw[thick] (3,-2.2)--(4.7,-3);
    \draw[thick] (3,-3.8)--(4.7,-3);
    \node at (3.6,-3) {$m$};
    \node at (-1,-3) {$i$};
    
    \draw[thick] (3,-7.2)--(3,-8.8);
    \draw[thick] (3,-7.2)--(4.7,-8);
    \draw[thick] (3,-8.8)--(4.7,-8);
    \node at (3.6,-8) {$n$};
    \node at (-1,-8) {$\bar{j}$};

\end{tikzpicture}
\end{array}\equiv|mn,i\bar{j},\pi\rangle,\ \pi\in\mathbb{S}^{k-1},\ m,n=1,2,\cdots,N
\end{equation}
where $i,\bar{j}$ label the position of the broken links. Let's act $\mathcal{P}$ on $|mn,i\bar{j},\pi\rangle$, and the five kinds of terms (labeled as $\boxed{1},\boxed{2},\boxed{3},\boxed{4},\boxed{5}$) are recorded in table~\ref{table:first excited state, acting P}.

Next, we act $\mathcal{X}+\bar{\mathcal{X}}$ on $|mn,i\bar{j},\pi\rangle$, and the four kinds of terms (labeled as $\boxed{2'},\boxed{3'},\boxed{5.1'},\boxed{5.2'}$) are recorded in table~\ref{table:first excited state, acting X}.

\begin{table}[t]
    \centering
    \begin{tabular}{c|c|c}
    \hline Label & Equation & \# \\
    \hline \boxed{2'} & $\adjincludegraphics[valign=c, width=0.1\textwidth]{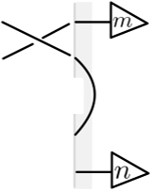}=\adjincludegraphics[valign=c, width=0.052\textwidth]{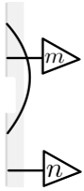}$ &  $k-1$ \\
    \hline \boxed{3'} & $\adjincludegraphics[valign=c, width=0.1\textwidth]{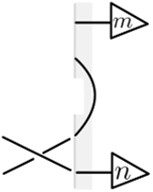}=\adjincludegraphics[valign=c, width=0.052\textwidth]{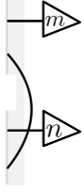}$ &  $k-1$ \\
    \hline \boxed{5.1'} & $\adjincludegraphics[valign=c, width=0.1\textwidth]{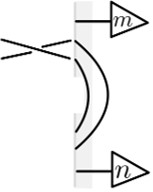}=\adjincludegraphics[valign=c, width=0.052\textwidth]{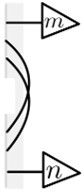}$ &  $\binom{k-1}{2}$ \\
    \hline \boxed{5.2'} & $\adjincludegraphics[valign=c, width=0.1\textwidth]{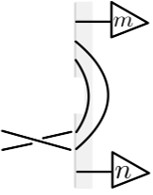}=\adjincludegraphics[valign=c, width=0.052\textwidth]{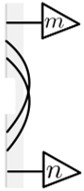}$ &  $\binom{k-1}{2}$ \\
    \hline
    \end{tabular}
    \caption{ $\left(\mathcal{X}+\bar{\mathcal{X}}\right)|mn,i\bar{j},\pi\rangle$ is grouped into four categories. Notice that the total number of terms is $2(k-1)+2\binom{k-1}{2}=2\binom{k}{2}$, which is correct.}
    \label{table:first excited state, acting X}
\end{table}

We notice that $\boxed{2}$ cancels $\boxed{2'}$, $\boxed{3}$ cancels $\boxed{3'}$, and $\boxed{5}$ cancels $\boxed{5.1'}+\boxed{5.2'}$. Therefore, the surviving terms are $\boxed{1}$ and $\boxed{4}$.

Grouping everything together, we obtain:
\begin{equation}
\begin{aligned}
\mathcal{L}_k|mn,i\bar{j},\pi\rangle&=-k|mn,i\bar{j},\pi\rangle+N^{-1}(k-1)\cdot N|mn,i\bar{j},\pi\rangle+N^{-1}\delta_{mn}\sum_{n'=1}^N|n'n',i\bar{j},\pi\rangle\\
&=-|mn,i\bar{j},\pi\rangle+N^{-1}\delta_{mn}|\pi'\rangle
\end{aligned}
\end{equation}
Notice that $|\pi'\rangle\equiv\sum_{n'=1}^N|n'n',i\bar{j},\pi\rangle, \pi'\in \mathbb{S}^{k}$ belongs to the ground state manifold. Therefore, we may consider a linear superposition to get rid of this term. Define $\alpha$ to be an $N \times N$ coefficient matrix:
\begin{equation}
|\alpha,i\bar{j},\pi\rangle\equiv\sum_{m,n=1}^N\alpha_{mn}|mn,i\bar{j},\pi\rangle
\end{equation}
Then we have:
\begin{equation}
\mathcal{L}_k|\alpha,i\bar{j},\pi\rangle=-|\alpha,i\bar{j},\pi\rangle+N^{-1}\tr[\alpha]\cdot|\pi'\rangle
\end{equation}
We see that by choosing $\alpha$ to be traceless, namely $\tr[\alpha]=0$, we obtain that $\mathcal{L}_k|\alpha,i\bar{j},\pi\rangle=-|\alpha,i\bar{j},\pi\rangle$. So we successfully construct the first excited states with eigenvalue $-1$.

Next, we aim to determine the degeneracy of the first excited states. There are $N^2-1$ choices for the traceless $N \times N$ matrix $\alpha$, $k^2$ choices for the positions of $m, n$ located at $i, \bar{j}=1,2,\cdots,k$, and $(k-1)!$ choices for $\pi \in \mathbb{S}^{k-1}$. Therefore, we may conclude that the degeneracy is $k \cdot k! \cdot (N^2-1)$.

We need to show that they are linearly independent. To do so, we analyze the inner product $\langle\alpha',i'\bar{j}',\pi'|\alpha,i\bar{j},\pi\rangle$ in the large $N$ limit. We first notice that when $(i',\bar{j}') \neq (i,\bar{j})$ or $\pi' \neq \pi$, the inner product is at most $O(N^{k-2})$. But for $(i',\bar{j}') = (i,\bar{j})$ and $\pi' = \pi$, the inner product is of order $O(N^{k-1})$. This shows the Gram matrix is almost block diagonal in the infinite $N$ limit. So, the states with $(i',\bar{j}') \neq (i,\bar{j})$ or $\pi' \neq \pi$ persist to be linearly independent in large but finite $N$ by continuity. Next, we consider the inner product within each block:
\begin{equation}
\langle\alpha',i\bar{j},\pi|\alpha,i\bar{j},\pi\rangle=N^{k-1}\tr[\alpha'^{\dagger}\alpha]
\end{equation}
This is a standard inner product for traceless matrices. So, there are indeed $N^2-1$ linearly independent choices of $\alpha$.

In summary, this shows that the degeneracy is indeed $k \cdot k! \cdot (N^2-1)$ at large but finite $N$. We expect that some null states will appear if $k$ is of order $O(N)$, which reduces degeneracy. This also happens in the ground state manifold, whose degeneracy starts to decrease from $k!$ when $k > N$~\cite{saad2019semiclassical}.

\subsubsection{Second excited states}
In this section, we study the family of second excited states of $\mathcal{L}_k$. The energy spectrum splits into $-2, -2 \pm 2N^{-1}$, and their corresponding degeneracy is summarized in table~\ref{table:summarize second excited states}.

Similarly, the second excited states are constructed by breaking two links of ground states. An example for $k=4$ is given by:

\begin{equation}
\begin{array}{c}
\begin{tikzpicture}[scale=0.35]
    \foreach \y in {1,2,3,4} {
        \coordinate (l\y) at (0, -\y);
    }
    \foreach \y in {5,6,7,8} {
        \coordinate (l\y) at (0, -\y-1);
    }

    \draw[thick,gray] (l1) .. controls +(right:3cm) and +(right:3cm) .. (l6);
    \draw[thick,gray] (l2) .. controls +(right:1cm) and +(right:1cm) .. (l5);
    \draw[thick,gray] (l3) .. controls +(right:2cm) and +(right:2cm) .. (l7);
    \draw[thick,gray] (l4) .. controls +(right:1cm) and +(right:1cm) .. (l8);

    \draw[lightgray] (l1)--(l4);
    \draw[lightgray] (l5)--(l8);

    \filldraw [lightgray] (0,-1) circle (4pt);
    \filldraw [lightgray] (0,-2) circle (4pt);
    \filldraw [lightgray] (0,-3) circle (4pt);
    \filldraw [lightgray] (0,-4) circle (4pt);
    \filldraw [lightgray] (0,-6) circle (4pt);
    \filldraw [lightgray] (0,-7) circle (4pt);
    \filldraw [lightgray] (0,-8) circle (4pt);
    \filldraw [lightgray] (0,-9) circle (4pt);

\end{tikzpicture}
\end{array}
\longrightarrow
\begin{array}{c}
\begin{tikzpicture}[scale=0.35]
    \foreach \y in {1,2,3,4} {
        \coordinate (l\y) at (0, -\y);
    }
    \foreach \y in {5,6,7,8} {
        \coordinate (l\y) at (0, -\y-1);
    }

    \draw[thick,gray] (l1) .. controls +(right:3cm) and +(right:3cm) .. (l6);

    \draw[thick,gray] (l3) .. controls +(right:2cm) and +(right:2cm) .. (l7);

    \draw[lightgray] (l1)--(l4);
    \draw[lightgray] (l5)--(l8);

    \filldraw [lightgray] (0,-1) circle (4pt);
    \filldraw [lightgray] (0,-2) circle (4pt);
    \filldraw [lightgray] (0,-3) circle (4pt);
    \filldraw [lightgray] (0,-4) circle (4pt);
    \filldraw [lightgray] (0,-6) circle (4pt);
    \filldraw [lightgray] (0,-7) circle (4pt);
    \filldraw [lightgray] (0,-8) circle (4pt);
    \filldraw [lightgray] (0,-9) circle (4pt);

    \draw[thick] (l2) -- (3,-2);
    \draw[thick] (l4) -- (3,-4);
    \draw[thick] (l5) -- (3,-6);
    \draw[thick] (l8) -- (3,-9);

    \draw[thick] (3,-3.7)--(3,-4.3);
    \draw[thick] (3,-3.7)--(3.8,-4);
    \draw[thick] (3,-4.3)--(3.8,-4);

    \draw[thick] (3,-1.7)--(3,-2.3);
    \draw[thick] (3,-1.7)--(3.8,-2);
    \draw[thick] (3,-2.3)--(3.8,-2);

    \draw[thick] (3,-5.7)--(3,-6.3);
    \draw[thick] (3,-5.7)--(3.8,-6);
    \draw[thick] (3,-6.3)--(3.8,-6);

    \draw[thick] (3,-8.7)--(3,-9.3);
    \draw[thick] (3,-8.7)--(3.8,-9);
    \draw[thick] (3,-9.3)--(3.8,-9);
    
    \node at (4.5,-2) {$m$};
    \node at (4.5,-4) {$n$};
    \node at (4.5,-6) {$p$};
    \node at (4.5,-9) {$q$};

    \node at (-1,-2) {$i_1$};
    \node at (-1,-4) {$i_2$};
    \node at (-1,-6) {$\bar{j}_1$};
    \node at (-1,-9) {$\bar{j}_2$};

\end{tikzpicture}
\end{array}\equiv|mnpq,i_1i_2\bar{j}_1\bar{j}_2,\pi\rangle,\ \pi\in\mathbb{S}^{k-2},\ m,n,p,q=1,2,\cdots,N
\end{equation}
where $i_1, i_2, \bar{j}_1, \bar{j}_2=1,2,\cdots,k$ label the position of broken links.

Let's act $\mathcal{L}_k$ on $|mnpq,i_1i_2\bar{j}_1\bar{j}_2,\pi\rangle$. From the experience in ground states and first excited states, we see that the following two kinds of terms in $\mathcal{P}$ will cancel those in $\mathcal{X} + \bar{\mathcal{X}}$: (1) exchange the ends of two unbroken links; (2) exchange the ends of one unbroken link and one broken link. So, the terms that are not canceled are the following four kinds of terms (labeled as $\boxed{1''}, \boxed{2''}, \boxed{3''}, \boxed{4''}$) recorded in table~\ref{table:second excited state}.

\begin{table}[t]
    \centering
    \begin{tabular}{c|c|c|c}
    \hline Label & Comes from where?& Equation & \# \\
    \hline \boxed{1''} & $\mathcal{P}$ & $\adjincludegraphics[valign=c, width=0.12\textwidth]{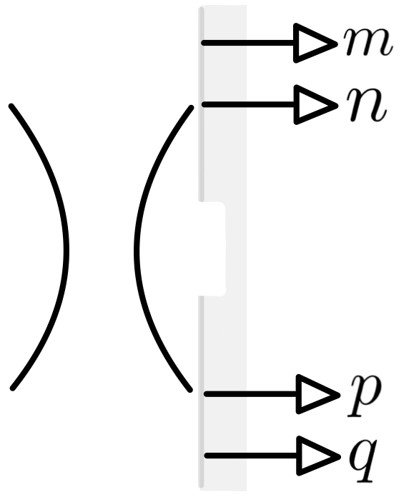}=\delta_{np}\cdot\adjincludegraphics[valign=c, width=0.062\textwidth]{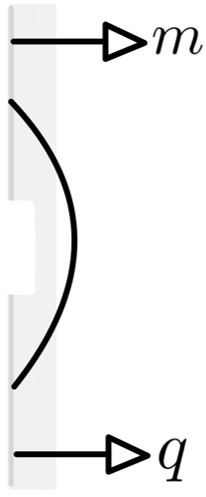}$ &  $4$ \\
    \hline \boxed{2''} & $\mathcal{P}$ & $\adjincludegraphics[valign=c, width=0.12\textwidth]{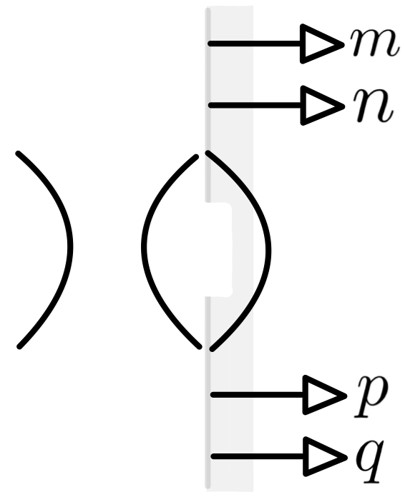}=N\cdot\adjincludegraphics[valign=c, width=0.062\textwidth]{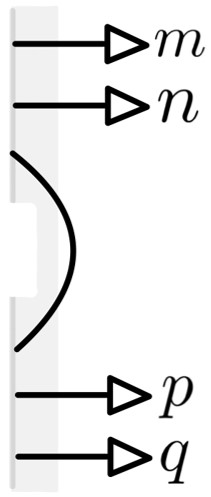}$ &  $k-2$ \\
    \hline \boxed{3''} & $\mathcal{X}$ & $\adjincludegraphics[valign=c, width=0.12\textwidth]{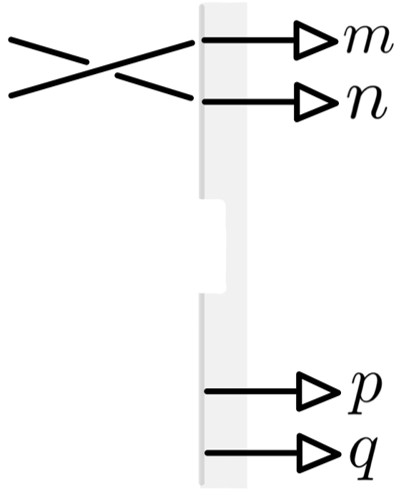}=\adjincludegraphics[valign=c, width=0.062\textwidth]{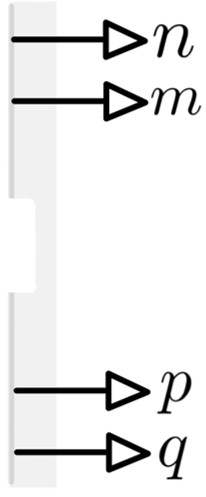}$ &  $1$ \\
    \hline \boxed{4''} & $\bar{\mathcal{X}}$ & $\adjincludegraphics[valign=c, width=0.12\textwidth]{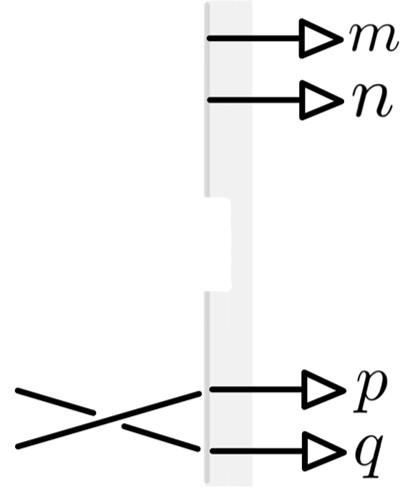}=\adjincludegraphics[valign=c, width=0.062\textwidth]{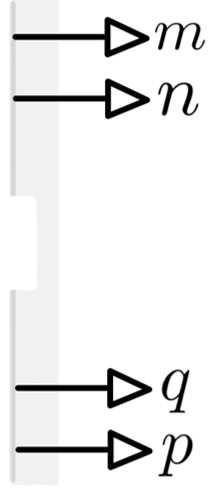}$ &  $1$ \\
    \hline
    \end{tabular}
    \caption{$(\mathcal{P} - \mathcal{X} - \bar{\mathcal{X}}) |mnpq, i_1 i_2 \bar{j}_1 \bar{j}_2, \pi\rangle$ is grouped into four kinds of terms.}
    \label{table:second excited state}
\end{table}

Putting everything together, we obtain:
\begin{equation}
\begin{aligned}
\mathcal{L}_k |mnpq, i_1 i_2 \bar{j}_1 \bar{j}_2, \pi\rangle & = -2 |mnpq, i_1 i_2 \bar{j}_1 \bar{j}_2, \pi\rangle - N^{-1} |nmpq, i_1 i_2 \bar{j}_1 \bar{j}_2, \pi\rangle - N^{-1} |mnqp, i_1 i_2 \bar{j}_1 \bar{j}_2, \pi\rangle \\
& \ \ \ + \delta_{mp} \sum_{m'=1}^N |m' nm' q, i_1 i_2 \bar{j}_1 \bar{j}_2, \pi\rangle + \delta_{mq} \sum_{m'=1}^N |m' npm', i_1 i_2 \bar{j}_1 \bar{j}_2, \pi\rangle \\
& \ \ \ + \delta_{np} \sum_{n'=1}^N |mn' n' q, i_1 i_2 \bar{j}_1 \bar{j}_2, \pi\rangle + \delta_{nq} \sum_{n'=1}^N |mn' pn', i_1 i_2 \bar{j}_1 \bar{j}_2, \pi\rangle
\end{aligned}
\end{equation}
The four terms on the second and third lines correspond to $\boxed{1''}$ in table~\ref{table:second excited state}. From the diagrams, these states have only one broken link, which belongs to the manifold of the first excited states. To construct eigenstates, we need these four terms to vanish. To achieve this, we define a $N \times N \times N \times N$ four-leg tensor $\alpha_{mnpq}$ and define the linear superposition state:
\begin{equation}
|\alpha, i_1 i_2 \bar{j}_1 \bar{j}_2, \pi\rangle = \sum_{m,n,p,q=1}^{N} \alpha_{mnpq} |mnpq, i_1 i_2 \bar{j}_1 \bar{j}_2, \pi\rangle
\end{equation}
Eliminating four terms from $\boxed{1''}$ is equivalent to imposing the following four linear constraints on $\alpha_{mnpq}$:
\begin{equation}
\sum_{m'=1}^N \alpha_{m' nm' q} = 0, \ \sum_{m'=1}^N \alpha_{m' npm'} = 0, \ \sum_{n'=1}^N \alpha_{mn' n' q} = 0, \ \sum_{n'=1}^N \alpha_{mn' pn'} = 0,
\end{equation}
Then, the equation is reduced to:
\begin{equation}
\begin{aligned}
\mathcal{L}_k |\alpha, i_1 i_2 \bar{j}_1 \bar{j}_2, \pi\rangle & = -2 |\alpha, i_1 i_2 \bar{j}_1 \bar{j}_2, \pi\rangle \\
& \ \ \ \ - N^{-1} \sum_{m,n,p,q=1}^N \alpha_{nmpq} |mnpq, i_1 i_2 \bar{j}_1 \bar{j}_2, \pi\rangle - N^{-1} \sum_{m,n,p,q=1}^N \alpha_{mnqp} |mnpq, i_1 i_2 \bar{j}_1 \bar{j}_2, \pi\rangle
\end{aligned}
\end{equation}
To make eigenstates, we need to further require that $\alpha$ is symmetric/anti-symmetric in its first two indices and symmetric/anti-symmetric in its last two indices. So, we obtain:
\begin{equation}
\label{eq:four kind of symmetry}
\left\{
\begin{aligned}
& \tikz[baseline=(char.base)]{\node[shape=circle,draw,inner sep=1pt] (char) {1};}:\ \mathcal{L}_k|\alpha,i_1i_2\bar{j}_1\bar{j}_2,\pi\rangle=-(2-2N^{-1})|\alpha,i_1i_2\bar{j}_1\bar{j}_2,\pi\rangle,\ \  \alpha_{mnpq}=-\alpha_{nmpq}=-\alpha_{mnqp} \\
& \tikz[baseline=(char.base)]{\node[shape=circle,draw,inner sep=1pt] (char) {2};}:\ \mathcal{L}_k|\alpha,i_1i_2\bar{j}_1\bar{j}_2,\pi\rangle=-(2+2N^{-1})|\alpha,i_1i_2\bar{j}_1\bar{j}_2,\pi\rangle, 
\ \ \alpha_{mnpq}=\alpha_{nmpq}=\alpha_{mnqp} \\
& \tikz[baseline=(char.base)]{\node[shape=circle,draw,inner sep=1pt] (char) {3};}:\ \mathcal{L}_k|\alpha,i_1i_2\bar{j}_1\bar{j}_2,\pi\rangle=-2|\alpha,i_1i_2\bar{j}_1\bar{j}_2,\pi\rangle, \ \ \ \ \ \ \ \ \ \ \ \ \ \ \ \alpha_{mnpq}=\alpha_{nmpq}=-\alpha_{mnqp} \\
& \tikz[baseline=(char.base)]{\node[shape=circle,draw,inner sep=1pt] (char) {4};}:\ \mathcal{L}_k|\alpha,i_1i_2\bar{j}_1\bar{j}_2,\pi\rangle=-2|\alpha,i_1i_2\bar{j}_1\bar{j}_2,\pi\rangle, \ \ \ \ \ \ \ \ \ \ \ \ \ \ \ \alpha_{mnpq}=-\alpha_{nmpq}=\alpha_{mnqp} \\
\end{aligned}
\right.
\end{equation}

Next, we need to count the degeneracy. Similar to the case of the first excited states, the inner product with different positions $(i_1 i_2 \bar{j}_1 \bar{j}_2)$ or different $\pi$ is at most of order $O(N^{k-3})$. Those states with the same position and permutation element have an inner product of order $O(N^{k-2})$. Therefore, states with different positions or different permutations are linearly independent in large but finite $N$. So, counting the position $(i_1 i_2 \bar{j}_1 \bar{j}_2)$ and permutation $\pi \in \mathbb{S}^{k-2}$ gives degeneracy $\binom{k}{2}^2 \times (k-2)! = \frac{k!}{4} k(k-1)$.

Then, we need to count the degeneracy of $\alpha$, satisfying symmetry conditions and four linear constraints. Notice that under such symmetry of indices, four constraints are satisfied if and only if any one of them is satisfied. We choose $\sum_{m'=1}^N \alpha_{m' nm' q} = 0$ as a representative. Since:
\begin{equation}
\langle \alpha', i_1 i_2 \bar{j}_1 \bar{j}_2, \pi | \alpha, i_1 i_2 \bar{j}_1 \bar{j}_2, \pi\rangle = N^{k-2} \sum_{m,n,p,q} \alpha'_{mnpq} \alpha_{mnpq} \equiv N^{k-2} (\alpha' | \alpha)
\end{equation}
where we define $|\alpha) = \sum_{m,n,p,q=1}^N \alpha_{mnpq} |mnpq)$ to be a state on an auxiliary Hilbert space that leaves the position $(i_1 i_2 \bar{j}_1 \bar{j_2})$ and permutation $\pi$ fixed and implicit.

Our strategy to construct the wave function $|\alpha)$ under symmetry and constraint is to start with an arbitrary $|\alpha_0)$, and then construct the projector $P$ that projects onto the subspace respecting symmetry and constraint. Then $|\alpha) = P|\alpha_0)$ is what we want. The number of linearly independent basis of $|\alpha)$ equals the rank of projector $P$, namely $\text{rank}(P) = \tr[P]$.

Before constructing projectors $P_{\circled{1}} \sim P_{\circled{4}}$ corresponding to four symmetry conditions in equation~(\ref{eq:four kind of symmetry}), we fix some notation on the auxiliary Hilbert space. The states $|\alpha)$ are diagrammatically represented by:
\begin{equation}
|\alpha) = \sum_{m,n,p,q=1}^N \alpha_{mnpq} |mnpq) = \sum_{m,n,p,q=1}^N \alpha_{mnpq} \adjincludegraphics[valign=c, width=0.07\textwidth]{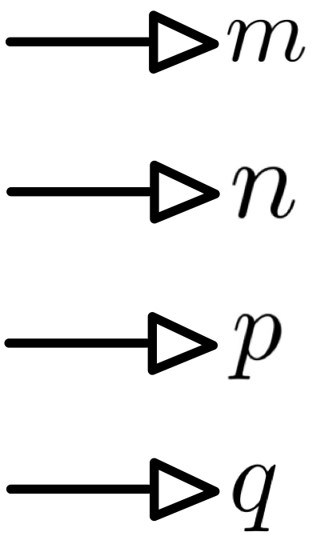}
\end{equation}
Define three kinds of projectors $P_s, P_a, P_0$ and a special state $|\psi)$:
\begin{equation}
P_s = \frac{1}{2} \left( \adjincludegraphics[valign=c, width=0.07\textwidth]{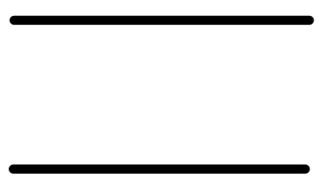} + \adjincludegraphics[valign=c, width=0.07\textwidth]{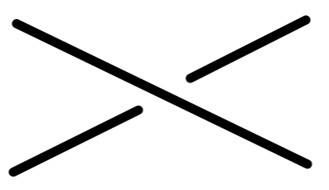} \right), \ P_a = \frac{1}{2} \left( \adjincludegraphics[valign=c, width=0.07\textwidth]{s_projector_2.jpg} - \adjincludegraphics[valign=c, width=0.07\textwidth]{s_projector_3.jpg} \right), \ P_0 = \adjincludegraphics[valign=c, width=0.06\textwidth]{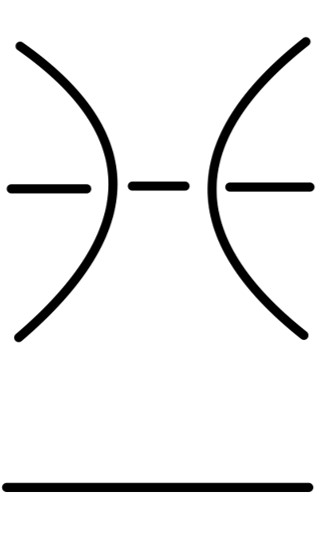}, \ |\psi) = \adjincludegraphics[valign=c, width=0.023\textwidth]{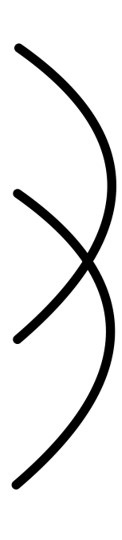}
\end{equation}
where $P_s$ is a symmetric projector and $P_a$ is an anti-symmetric projector. Then, $P_{\circled{1}} \sim P_{\circled{4}}$ are given by:
\begin{equation}
\begin{aligned}
& P_{\circled{1}} = P_a \otimes P_a - \frac{4}{N-2} (P_a \otimes P_a) P_0 (P_a \otimes P_a) + \frac{2}{(N-1)(N-2)} (P_a \otimes P_a) |\psi) (\psi | (P_a \otimes P_a) \\
& P_{\circled{2}} = P_s \otimes P_s - \frac{4}{N+2} (P_s \otimes P_s) P_0 (P_s \otimes P_s) + \frac{2}{(N+1)(N+2)} (P_s \otimes P_s) |\psi) (\psi | (P_s \otimes P_s) \\
& P_{\circled{3}} = P_s \otimes P_a - \frac{4}{N} (P_s \otimes P_a) P_0 (P_s \otimes P_a) \\
& P_{\circled{4}} = P_a \otimes P_s - \frac{4}{N} (P_a \otimes P_s) P_0 (P_a \otimes P_s)
\end{aligned}
\end{equation}
One can check that $P_{\circled{i}} |\alpha_0)$ satisfies the symmetry condition and constraint corresponding to $\circled{i}$. Also, one can check that $P_{\circled{i}} P_{\circled{j}} = \delta_{ij} P_{\circled{i}}$, which is indeed a properly orthonormalized projector. Picking up $\frac{k!}{4} k(k-1)$ factor from the degeneracy of position and permutation, then by evaluating $\tr[P_{\circled{i}}]$, we can obtain the full degeneracy, which we summarize in table~\ref{table:summarize second excited states}.

\begin{table}[h]
    \centering
    \begin{tabular}{c|c}
    \hline
       Eigenvalue of $\mathcal{L}_k$ & Degeneracy \\
    \hline
       $-(2-2N^{-1})$ & $\frac{k!}{4} k(k-1) \times \tr[P_{\circled{1}}] = \frac{k!}{4} k(k-1) \times \frac{1}{4} N^2 (N+1) (N-3)$ \\
    \hline
       $-2$ & $\frac{k!}{4} k(k-1) \times (\tr[P_{\circled{3}}] + \tr[P_{\circled{4}}]) = \frac{k!}{4} k(k-1) \times \frac{1}{2} (N^2-4) (N^2-1)$ \\
    \hline
       $-(2+2N^{-1})$ & $\frac{k!}{4} k(k-1) \times \tr[P_{\circled{2}}] = \frac{k!}{4} k(k-1) \times \frac{1}{4} N^2 (N-1) (N+3)$ \\
    \hline
    \end{tabular}
    \caption{Summary of energy and degeneracy for second excited states}
    \label{table:summarize second excited states}
\end{table}

\subsubsection{General excited states in the infinite $N$ limit}
\label{sec:general high excited state in infinite $N$ limit}
For general highly excited states, it is challenging to exactly solve the wave function, energy, and degeneracy. However, in the infinite $N$ limit, things simplify.

Consider the $p^{\text{th}}$ excited states ($p = 0, 1, 2, \ldots, k$), whose wave function is similarly constructed by breaking $p$-links on top of ground states, and then gluing computational basis to broken ends. These states are labeled by $|m_1 \ldots m_p n_1 \ldots n_p, i_1 \ldots i_p \bar{j}_1 \ldots \bar{j}_p, \pi\rangle, \pi \in \mathbb{S}^{k-p}$. Action of $N^{-1} (\mathcal{P} - \mathcal{X} - \bar{\mathcal{X}}) \subset \mathcal{L}_k$ on it will generate four kinds of terms analogous to $\boxed{1''}, \boxed{2''}, \boxed{3''}, \boxed{4''}$ in table~\ref{table:second excited state}. In the infinite $N$ limit, $\boxed{1''}, \boxed{3''}, \boxed{4''}$ are suppressed by $N^{-1}$ while $\boxed{2''}$ are not. Therefore, $\mathcal{L}_k \approx -k + \boxed{2''}$:
\begin{equation}
\begin{aligned}
\lim_{N\rightarrow\infty} \mathcal{L}_k |m_1 \ldots m_p n_1 \ldots n_p, i_1 \ldots i_p \bar{j}_1 \ldots \bar{j}_p, \pi\rangle & = \left[ -k + N^{-1} (k-p) N \right] |m_1 \ldots m_p n_1 \ldots n_p, i_1 \ldots i_p \bar{j}_1 \ldots \bar{j}_p, \pi\rangle \\
& = -p |m_1 \ldots m_p n_1 \ldots n_p, i_1 \ldots i_p \bar{j}_1 \ldots \bar{j}_p, \pi\rangle
\end{aligned}
\end{equation}
So, they are eigenstates with energy $-p$. The degeneracy of these states is counted to be $\binom{k}{p}^2 (k-p)! N^{2p}$.

\subsection{Approaching $k$-design}
\label{section:calculating frame potential}
Given an ensemble $\mathcal{E}$ of unitaries acting on Hilbert space $\mathcal{H}$ with dimension $N$, we can ask how close $\mathcal{E}$ is to the Haar ensemble. The distance between two ensembles can be measured by comparing their moments. Let $X$ be an operator on $k$-copies of Hilbert space: $\mathcal{H}^{\otimes k}$. Then define the following \textit{moment map} channel~\cite{Jian:2022pvj}:
\begin{equation}
\Phi^{(k)}_{\mathcal{E}} : X \longrightarrow \E_{U \in \mathcal{E}} \left[ \left( U^{\otimes k} \right) X \left( U^{\dagger \otimes k} \right) \right]
\end{equation}
Then we say that $\mathcal{E}$ is an $\varepsilon$-approximated $k$-design if the distance between two channels $\Phi^{(k)}_{\mathcal{E}}, \Phi^{(k)}_{\text{Haar}}$ is smaller than $\varepsilon$ measured in the diamond norm~\cite{Jian:2022pvj}:
\begin{equation}
\lVert \Phi^{(k)}_{\mathcal{E}} - \Phi^{(k)}_{\text{Haar}} \rVert_{\diamond} < \varepsilon
\end{equation}
Since the diamond norm is usually hard to calculate, we may use the frame potential:
\begin{equation}
F^{(k)}_{\mathcal{E}} \equiv \E_{U \in \mathcal{E}} \E_{V \in \mathcal{E}} \left| \tr \left( UV^{\dagger} \right) \right|^{2k}
\end{equation}
to bound the diamond norm via the following inequality~\cite{Jian:2022pvj}:
\begin{equation}
\lVert \Phi^{(k)}_{\mathcal{E}} - \Phi^{(k)}_{\text{Haar}} \rVert_{\diamond}^2 \leq N^{2k} (F^{(k)}_{\mathcal{E}} - F^{(k)}_{\text{Haar}})
\end{equation}
The frame potential for the Haar ensemble is given by $F^{(k)}_{\text{Haar}} = k!$. In our BGUE case, we find that:
\begin{equation}
\begin{aligned}
F^{(k)}_\BGUE(t) & = \E_{U_t, V_t} \tr \left[ \left( U_t^{\otimes k} \otimes U_t^{* \otimes k} \right) \cdot \left( V_t^{\otimes k} \otimes V_t^{* \otimes k} \right)^{\dagger} \right] \\
& = \tr \left[ \mathcal{U}_k(t) \cdot \mathcal{U}_k(t)^{\dagger} \right] = \tr \left[ \mathcal{U}_k(2t) \right] = \tr \left[ e^{2t \cdot \mathcal{L}_k} \right]
\end{aligned}
\end{equation}
where we use the fact that $\mathcal{L}_k = \mathcal{L}_k^{\dagger}$ and $\mathcal{U}_k(t) \mathcal{U}_k(t) = \mathcal{U}_k(2t)$. We see that the frame potential only depends on the spectrum and degeneracy of $\mathcal{L}_k$. Therefore, using the results of ground states, first excited states, and second excited states, we can approximate the frame potential in the long-time limit:
\begin{equation}
\begin{aligned}
F^{(k)}_\BGUE(t) = k! & + k! k (N^2-1) e^{-2t} + k! \frac{k(k-1)}{16} N^2 (N+1) (N-3) e^{-(4-4N^{-1})t} \\
& + k! \frac{k(k-1)}{8} (N^2-4) (N^2-1) e^{-4t} + k! \frac{k(k-1)}{16} N^2 (N-1) (N+3) e^{-(4+4N^{-1})t} \\
& + O(e^{-6t})
\end{aligned}
\label{eq:frame potential for k-replica}
\end{equation}
Neglecting second excited states, we obtain the time for BGUE to approach an $\varepsilon$-approximated $k$-design is:
\begin{equation}
t_k \geq \frac{1}{2} \left[ \log \varepsilon^{-2} + k \log N + \log (N^2-1) + \log k! + \log k \right]
\label{eq:time to approach k-design}
\end{equation}
The linear-in-$k$ scaling is consistent with the Brownian SYK result~\cite{Jian:2022pvj}.

One can also use the result in section~\ref{sec:general high excited state in infinite $N$ limit} to calculate the frame potential in a certain limit: scaling $N \rightarrow \infty, t \rightarrow +\infty$ while keeping $Ne^{-t} \rightarrow e^{-\tau}$ finite:
\begin{equation}
F_\BGUE^{(k)}(t) \longrightarrow \sum_{p=0}^k \binom{k}{p}^2 (k-p)! N^{2p} \cdot e^{-2pt} = k! \sum_{p=0}^k \binom{k}{p} \frac{1}{p!} \left( N^2 e^{-2t} \right)^p
\end{equation}
Using Laguerre polynomial $L_k(x) = \sum_{p=0}^k \binom{k}{p} \frac{1}{p!} (-x)^p$, we obtain that:
\begin{equation}
F_\BGUE^{(k)}(t) \longrightarrow F^{(k)}_{\text{Haar}} \times L_k(-e^{-2\tau})
\end{equation}

\section{Application in classical shadow tomography}
\label{section: applications to shadow tomography}

\subsection{Brief review of classical shadow tomography}
The method of classical shadow tomography aims to predict many observables from just a few measurements. The central spirit of this method is: \textit{First do measurement, and then ask questions}~\cite{Huang2020,Elben2023,PhysRevLett.127.030503,PhysRevLett.126.190505}.

An illuminating example is: given an unknown single qubit state $\rho$, we want to predict the expectation values of all three Pauli observables $\tr[\rho X],\tr[\rho Y],\tr[\rho Z]$ up to error $\varepsilon$ by performing as few measurements as possible. The observation is that since $X,Y,Z$ do not commute, we cannot measure them simultaneously. In other words, if we measure $X$ to high precision with an error smaller than $\varepsilon$ (e.g., we perform measurements in the $X$-eigenbasis), then the error of predicting $\tr[\rho Z]$ would be much larger than $\varepsilon$. Therefore, instead of measuring in the eigenbasis of $X,Y,Z$ many times, we measure in some random directions a few times, so that each measurement can moderately predict all three $\tr[\rho X],\tr[\rho Y],\tr[\rho Z]$.

Now, we are ready to formulate the task and method of classical shadow tomography~\cite{Huang2020}.
Consider we are handed an experimental oracle that can produce an unknown state $\rho$ on an $n$-qubit Hilbert space. We want to perform some measurements on $\rho$, and then predict the expectation value $\tr[\rho O_i]$ of a set of $M$-observables $\{O_1,O_2,...,O_M\}$. The idea is similar: we perform some random measurements on $\rho$. This is achieved by applying a random unitary $U\in\mathcal{E}$ ($\mathcal{E}$ is a pre-designed ensemble of unitaries) on $\rho$, and then perform measurements on the computational basis, with the measurement result being $b\in\{0,1\}^{\otimes n}$. We perform such procedures for $K$ times, each time recording a pair $(U_i,b_i)$.

Define $\mathcal{M}$ to be the measurement channel:
\begin{equation}
\mathcal{M}[\rho]\equiv\sigma\equiv\E_{\hat{\sigma}}\hat{\sigma}\equiv\E_{U}\sum_b\tr[\rho\hat{\sigma}_{U,b}]\hat{\sigma}_{U,b},\ \hat{\sigma}_{U,b}\equiv U^{\dagger}|b\rangle\langle b|U
\end{equation}
For a tomographically complete channel~\cite{Huang2020}, $\mathcal{M}$ is invertible. Therefore, we obtain:
\begin{equation}
\rho=\mathcal{M}^{-1}[\sigma]=\E_{\hat{\sigma}}\mathcal{M}^{-1}[\hat{\sigma}]
\end{equation}
In practice, we obtain $K$ realizations of $\hat{\sigma}_i\equiv U^{\dagger}_i|b_i\rangle\langle b_i|U_i$, therefore an unbiased estimation of $\rho$ is by replacing $\E_{\hat{\sigma}}$ by a finite sum:
\begin{equation}
\rho_{\text{est}}=\frac{1}{K}\sum_{i=1}^K\mathcal{M}^{-1}[\hat{\sigma}_i]
\end{equation}
Then we can use $\rho_{\text{est}}$ to calculate $\tr[\rho_{\text{est}}O]$, which is an unbiased estimator of $\tr[\rho O]$:
\begin{equation}
o\equiv\tr[\rho O]=\E_{\hat{\sigma}}\tr(O\mathcal{M}^{-1}[\hat{\sigma}])\equiv\E_{\hat{o}}\hat{o}, \ o_{\text{est}}=\frac{1}{K}\sum_{i=1}^K\hat{o}_i
\end{equation}
The performance or accuracy of $o_{\text{est}}$ is related to the variance of the classical random variable $\hat{o}$, which is bounded by:
\begin{equation}
\text{Var}[\hat{o}]\leq\E_{\hat{o}}[\hat{o}^2]=\E_{U}\sum_{b}\langle b|U\rho U^{\dagger}|b\rangle\cdot|\langle b|U\mathcal{M}^{-1}[O]U^{\dagger}|b\rangle|^2
\end{equation}
Since one wants to define a quantity that quantifies the variance for generic $\rho$, which only depends on $O$, not $\rho$, we can further maximize over $\rho$ or take an average over $\rho$. Finally, we obtain a max-version or a mean-version of the shadow norm:
\begin{equation}
\begin{aligned}
\text{max-version: }&\lVert O\rVert^2_{\text{sh}}\equiv\max_{\rho}\E_{U}\sum_{b}\langle b|U\rho U^{\dagger}|b\rangle\cdot|\langle b|U\mathcal{M}^{-1}[O]U^{\dagger}|b\rangle|^2\\
\text{mean-version: }&\lVert O\rVert^2_{\text{sh}}\equiv\text{mean}_{\rho}\E_{U}\sum_{b}\langle b|U\rho U^{\dagger}|b\rangle\cdot|\langle b|U\mathcal{M}^{-1}[O]U^{\dagger}|b\rangle|^2\\
\end{aligned}
\label{eq:definition of shadow norm}
\end{equation}

We notice that the measurement channel $\mathcal{M}$ and therefore the shadow norm depends on the choice of random unitary ensemble $U\in\mathcal{E}$. Therefore, for a given type of interested $O$, we need to design $\mathcal{E}$ to make the shadow norm as small as possible, which in turn decreases the number of measurements needed~\cite{Huang2020}. For example, if $O$ is a Pauli string with length $\ell$, and $\mathcal{E}$ is an ensemble of shallow circuits with depth $d$ (one can imagine that for instance every $U\in \mathcal{E}$ is a brick-wall circuit with depth-$d$, with every local 2-qubit-gate being Haar-random), then there exists an optimal depth $d_*(\ell)$ that minimizes the shadow norm~\cite{PhysRevLett.130.230403,PhysRevResearch.5.023027}. See~\cite{PRXQuantum.5.020304,PhysRevLett.130.230403,PhysRevResearch.4.013054,PhysRevResearch.5.023027,Akhtar2023scalableflexible,PhysRevB.109.094209,akhtar2024dualunitary} for more schemes of designing $\mathcal{E}$ and~\cite{chen2021hierarchy,9719827,Aharonov2022} for information-theoretical upper/lower bounds that assess the performance of the shadow tomography method.

\subsection{Classical shadow tomography using BGUE}
In this section, we apply BGUE to the task of classical shadow tomography~\cite{Huang2020}. We derive explicit analytical results on shadow norm $\lVert O\rVert^2_{\text{sh}}(t)$ as a function of $t$. We observe that an optimal time $t_*(x_o)$ exists as a function of the off-diagonal portion $x_o$ of $O$, which minimizes the shadow norm. This phenomenon is analogous to the optimal circuit depth $d_*(\ell)$ in the shallow-circuit protocol~\cite{PhysRevLett.130.230403,PhysRevResearch.5.023027}.

Let's start by calculating the measurement channel. The measurement channel acting on the density matrix is defined as:
\begin{equation}
\mathcal{M}_t[\rho]=\E_{U_t}\sum_{b}\tr[\rho\hat{\sigma}_{U_t,b}]\hat{\sigma}_{U_t,b}=\E_{U_t}\sum_{b}\langle b|U_t\rho U_t^{\dagger}|b\rangle\cdot U^{\dagger}_t|b\rangle\langle b| U_t
\end{equation}
This can be evaluated exactly using our two-replica result in section~\ref{section:1/2 replica}, using the diagrams in equation~(\ref{eq:definition of A_i}). The result is given by:
\begin{equation}
\mathcal{M}_t[\rho]=N^{-1}\tr[\rho]\cdot\mathbb{I}+N^{-1}\beta_o(t)^{-1}\rho_o+N^{-1}\beta_d(t)^{-1}\rho_d
\end{equation}
where $\rho_o$ is the off-diagonal part of $\rho$ and $\rho_d$ is the traceless diagonal part of $\rho$.
The coefficients are given by:
\begin{equation}
\begin{aligned}
N^{-1}\beta_o(t)^{-1}&=2f_2(t)+4f_4(t)+Nf_5(t)+Nf_7(t)+2f_8(t)=\frac{1-e^{-(2+2N^{-1})t}}{N+1}\\
N^{-1}\beta_d(t)^{-1}&=f_1(t)+2f_2(t)+2f_3(t)+4f_4(t)+Nf_5(t)+f_6(t)+Nf_7(t)+2f_8(t)=\frac{1+e^{-(2+2N^{-1})t}}{N+1}\\
\end{aligned}
\end{equation}
where $f_i(t)$ is defined in equation~(\ref{eq:define f_i(t)}). The inverse channel can be readily calculated:
\begin{equation}
\mathcal{M}_t^{-1}[O]=N^{-1}\tr[O]\cdot\mathbb{I}+\beta_o(t)O_o+\beta_d(t)O_d
\end{equation}
\paragraph{Shadow Norm. }Similarly, $O_o,O_d$ are the off-diagonal and traceless diagonal part of $O$, respectively. Next, we calculate the (mean-version) shadow norm for a traceless Hermitian operator $O$:
\begin{equation}
\begin{aligned}
\lVert O\rVert_{\text{sh}}^2(t)&\equiv\text{mean}_{\rho}\E_{U_t}\sum_{b}\langle b|U_t\rho U_t^{\dagger}|b\rangle\cdot|\langle b|U_t\mathcal{M}^{-1}_t[O]U^{\dagger}_t|b\rangle|^2\\
&=N^{-1}\E_{U_t}\sum_{b}|\langle b|U_t\mathcal{M}^{-1}_t[O]U^{\dagger}_t|b\rangle|^2\\
&=N^{-1}\tr\left(\mathcal{M}_t^{-1}[O]\mathcal{M}_t\left[\mathcal{M}_t^{-1}[O]\right]\right)\\
&=N^{-1}\tr\left(\mathcal{M}_t^{-1}[O]O\right)\\
&=\beta_o(t)\tr\left(O_o^2\right)+\beta_d(t)\tr\left(O_d^2\right)+(\beta_o(t)+\beta_d(t))\tr\left(O_oO_d\right)\\
&=\beta_o(t)\tr\left(O_o^2\right)+\beta_d(t)\tr\left(O_d^2\right)\\
\end{aligned}
\end{equation}
where we used the fact that $\tr\left(O_oO_d\right)=0$. Notice that $\tr\left(O^2\right)=\tr\left(O_o^2\right)+\tr\left(O_d^2\right)$. So, we may define $x_o,x_d$ to be the portion of the off-diagonal part and traceless diagonal part inside $O$: $x_o\equiv\tr\left(O_o^2\right)\big/\tr\left(O^2\right)$, $x_d=\tr\left(O_d^2\right)\big/\tr\left(O^2\right)$, with the constraint $x_o+x_d=1,\ x_o,x_d\in[0,1]$. Thus, the shadow norm in units of the two-norm is given by:
\begin{equation}
\lVert O\rVert_{\text{sh}}^2(t)\big/\lVert O\rVert_2^2=\beta_o(t)x_o+\beta_d(t)x_d=\frac{N+1}{N}\left(\frac{x_o}{1-e^{-(2+2N^{-1})t}}+\frac{x_d}{1+e^{-(2+2N^{-1})t}}\right)
\label{eq: shadow norm}
\end{equation}

\begin{figure}[t]
    \centering
    \includegraphics[width=0.49\textwidth]{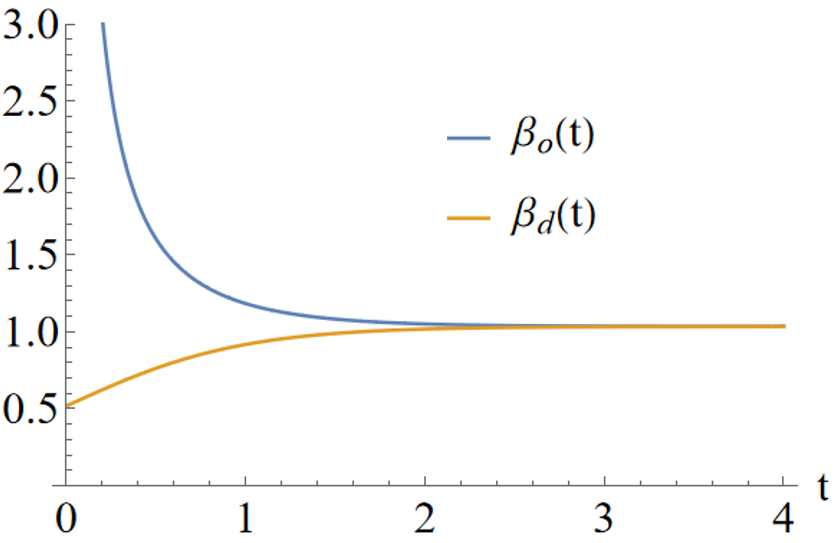}
    \includegraphics[width=0.49\textwidth]{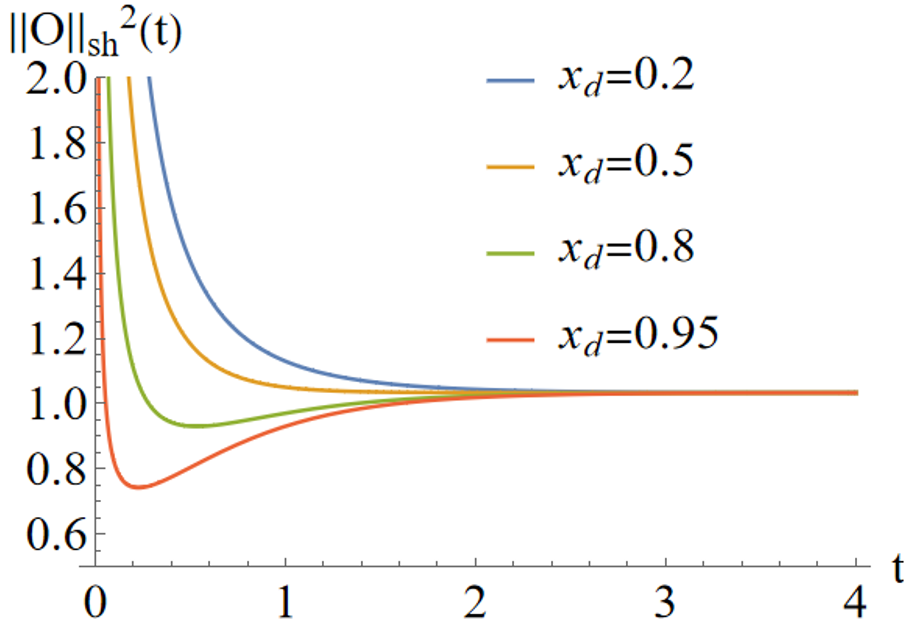}    
    \caption{\textbf{Left:} Shadow norm for purely diagonal or purely off-diagonal operator. \textbf{Right:} Shadow norm~(\ref{eq: shadow norm}) with mixing diagonal and off-diagonal entry. For both panels, we choose $N=30$, and the shadow norm is measured in unit of $\lVert O\rVert_2^2$.}
    \label{fig:shadow norm}
\end{figure}

We first analyze the shadow norm of a purely diagonal ($x_d=1$) or purely off-diagonal operator ($x_o=1$). We see that $\beta_d(t)$ increases with time. This means that the scrambling will decrease the efficiency of predicting the diagonal operator. On the other hand, $\beta_o(t)$ decreases with time. This means that the scrambling will improve the efficiency of predicting the off-diagonal operator. We notice that $\beta_o(t=0)=+\infty$, which means in the initial time, we cannot predict the off-diagonal operator at all. This is reasonable since without $U_t$ we are just measuring the computational basis, which only knows about the diagonal entries of the operator. We also notice that as $t\rightarrow\infty$, $\beta_o=\beta_d=\frac{N+1}{N}$, approaching the global Haar random result~\cite{Huang2020}. This is reasonable since Haar randomness is basis independent; therefore, diagonal and off-diagonal should have the same shadow norm. A plot of these phenomena can be found in figure~\ref{fig:shadow norm}.

\paragraph{Optimal Time. }Now, we consider $O$ has both $O_o$ and $O_d$ components. It turns out that when $x_d>x_o$, there is an optimal time $t_*$ such that $||O||_{\text{sh}}^2(t)$ is minimal:
\begin{equation}
t_*(x_o)=
\left\{
\begin{aligned}
&\frac{1}{2+2N^{-1}}\log\left(\frac{1+\sqrt{1-(x_d-x_o)^2}}{x_d-x_o}\right),\ & x_d>x_o\\
&+\infty,\ &x_d\leq x_o 
\end{aligned}
\right.
\label{eq:optimal time of shadow norm}
\end{equation}
The minimal shadow norm is given by:
\begin{equation}
\lVert O\rVert_{\text{sh}}^2(t_*)\big/\lVert O\rVert_2^2=
\left\{
\begin{aligned}
&\frac{N+1}{N}\cdot\frac{1}{2}\left[1+\sqrt{1-(x_d-x_o)^2}\right],\ & x_d>x_o\\
&\frac{N+1}{N},\ &x_d\leq x_o 
\end{aligned}
\right.
\end{equation}
This phenomenon is observed in numerical simulations in figure~\ref{fig:shadow norm}. The existence of an optimal time $t_*(x_o)$ as a function of the off-diagonal portion $x_o$ is analogous to the optimal circuit depth $d_*(\ell)$ as a function of the Pauli-string's length $\ell$ in shallow-circuit classical shadow tomography~\cite{PhysRevLett.130.230403,PhysRevResearch.5.023027}.

\section{Conclusion}

In this study, we have constructed and thoroughly analyzed the Brownian generalization of the Gaussian Unitary Ensemble (BGUE). Our investigation encompassed both the non-equilibrium dynamics, complexity and test of applications in classical shadow tomography, yielding significant insights into its behavior and potential applications in broaden quantum information tasks.

We began the study of non-equilibrium dynamics in BGUE by deriving explicit analytical expressions for various one-replica and two-replica variables at finite \(N\) and \(t\). These variables include the spectral form factor, its fluctuation, the two-point function, the fluctuations of the two-point function, out-of-time-order correlators (OTOC), the second Rényi entropy, and the frame potential for unitary 2-designs. 

Our results revealed that BGUE exhibits hyperfast scrambling and the emergence of tomperature. Additionally, we identified replica-wormhole-like contributions, leading to a non-decaying two-point function as \(t \to +\infty\) and non-vanishing fluctuations of the two-point function with zero mean value.

We further explored the complexity of BGUE, determining the time required for the ensemble to approximate the Haar ensemble in terms of \(k\)-design. Our findings indicate that the timescale is linear in \(k \log N\), consistent with previous studies on the Brownian Sachdev-Ye-Kitaev (SYK) model. In our investigation of the low-energy eigen-wavefunctions, spectrum, and degeneracy of the effective imaginary-time evolution operator on \(2k\)-replicated contours, we derived long-time results for the frame potential, confirming that the time required for BGUE to reach \(k\)-design is linear in \(k\).

In the final section, we applied the BGUE model to classical shadow tomography. We derived analytical results for the shadow norm and identified an optimal time that minimizes the shadow norm, analogous to the optimal circuit depth in shallow-circuit shadow tomography.

There are some possible interesting questions based on our study of BGUE:
\begin{itemize}
\item[1.]\textit{Exact $\mU_3(t)$ for three-replica observables.} Many interesting observables need three replicas. One relevant example is the max-version shadow norm defined in~(\ref{eq:definition of shadow norm}).
\item[2.]\textit{Application in other quantum information tasks.} For example, we can study the quantum causal influence (QCI)~\cite{Cotler_2019QCI}, spacetime entanglement entropy in super-density operator formalism~\cite{Cotler_2018Superdensity}, and channel capacity~\cite{Hosur_2016}. We are currently exploring these directions.
\item[3.]\textit{Adding noise and measurement.} We can add decoherence and measurement to the scrambling of BGUE, which can be used to study Measurement Induced Phase Transition or Noise Induced Phase Transition extensively studied in recent literature.
\item[4.]\textit{Brownian generalization of other ensemble.} One direct generalization is consider different symmetry class, like Brownian Gaussian Orthogonal Ensemble (BGOE) and Brownian Sympletic Ensemble (BGSE).
\end{itemize}

In conclusion, our study provides a comprehensive understanding of the non-equilibrium dynamics and complexity of BGUE, demonstrating its  scrambling behavior and potential for its applications in quantum information tasks such as classical shadow tomography. The analytical techniques and results presented here lay a foundation for future explorations of Brownian random matrix models and their applications in quantum physics and beyond.

\acknowledgments
I would like to thank Wanda Hou, Xiao-Liang Qi, Douglas Stanford, Hong-Yi Wang, Jinzhao Wang, Jiuci Xu, Shunyu Yao, Yi-Zhuang You, and Yuzhen Zhang for helpful discussion. I would especially thank Yuzhen Zhang for stimulating conversations on frame potential. I would also thank numerous useful discussions with Hong-Yi Wang on classical shadow tomography. I thank professor Xiao-Liang Qi's support by National Science Foundation under grant No.2111998 and the Simons Foundation.

\appendix
\section{Subset of High Energy Spectrum}
\label{section: subset of high energy spectrum}
In this appendix, we aim to gain a deeper understanding of the generic high-energy spectrum of $\mathcal{L}_k$. Although deriving the general formula for the spectrum and degeneracy by breaking more links on ground states is challenging, we demonstrate a special case where $\{-\left[p+N^{-1}p(p-1)\right] | p=0,1,\cdots,k\}$ is a subset of the spectrum of $\mathcal{L}_k$, using a different method. To achieve this, we work in an operator basis instead of a state basis.

Consider $B^{(k)}_p$ as an \textit{operator} on the Hilbert space $\mathbb{C}^{\otimes(2k N)}$. $B_p^{(k)}$ is defined by the equal-weight summation of all diagrams with: (1) $p$-lines connecting upper left and upper right; (2) $p$-lines connecting lower left and lower right; (3) $(k-p)$-lines connecting upper left and lower left; and (4) $(k-p)$-lines connecting upper right and lower right. The number of terms in the summand of $B_p^{(k)}$ is:
\begin{equation}
\#\big(B_p^{(k)}\big)=\binom{k}{p}^4[(k-p)!]^2[p!]^2
\end{equation}
\begin{table}[t]
    \centering
    \begin{tabular}{c|c|c|c|c}
    \hline Label & Source & Equation & Count & Mapping of diagrams \\
    \hline \boxed{1'''} & $\mathcal{P}$ & $\adjincludegraphics[valign=c, width=0.16\textwidth]{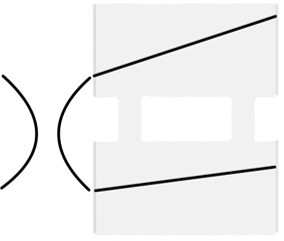}=\cdot\adjincludegraphics[valign=c, width=0.11\textwidth]{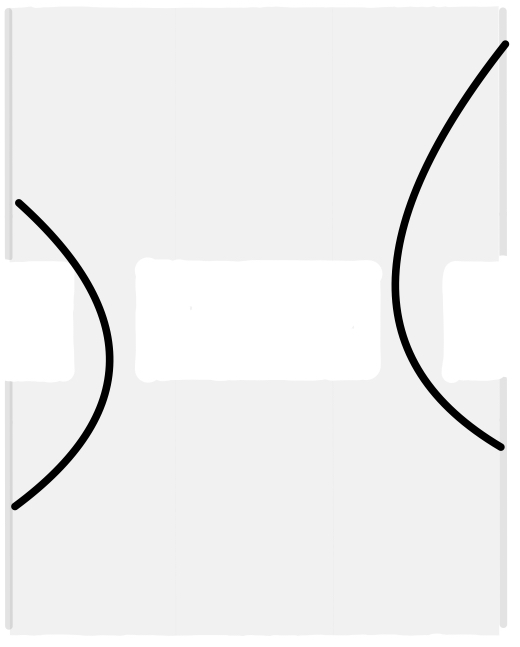}$ &  $p^2$ & $B_p^{(k)}\longrightarrow B_{p-1}^{(k)}$\\
    \hline \boxed{2'''} & $\mathcal{P}$ & $\adjincludegraphics[valign=c, width=0.16\textwidth]{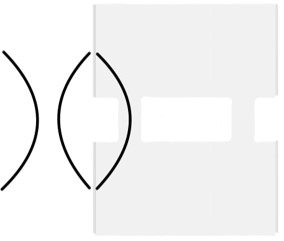}=N\cdot\adjincludegraphics[valign=c, width=0.108\textwidth]{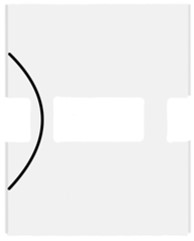}$ &  $k-p$ & $B_p^{(k)}\longrightarrow B_{p}^{(k)}$\\
    \hline \boxed{3'''} & $\mathcal{X}$ & $\adjincludegraphics[valign=c, width=0.16\textwidth]{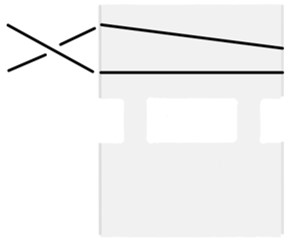}=\adjincludegraphics[valign=c, width=0.108\textwidth]{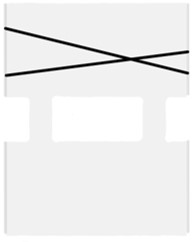}$ &  $\binom{p}{2}$ & $B_p^{(k)}\longrightarrow B_{p}^{(k)}$\\
    \hline \boxed{4'''} & $\bar{\mathcal{X}}$ & $\adjincludegraphics[valign=c, width=0.16\textwidth]{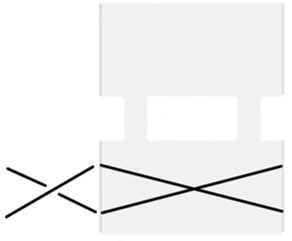}=\adjincludegraphics[valign=c, width=0.108\textwidth]{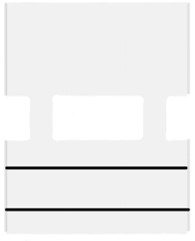}$ &  $\binom{p}{2}$ & $B_p^{(k)}\longrightarrow B_{p}^{(k)}$\\
    \hline
    \end{tabular}
    \caption{Action of $\mathcal{L}_k$ on one term in the summand of $B_p^{(k)}$. This table directly corresponds to table~\ref{table:second excited state} row by row.}
    \label{table:subset of high excited states}
\end{table}
An example of a diagram within the summand of $B_3^{(6)}$ would be:
\begin{equation}
B_{3}^{(6)}\supset \adjincludegraphics[valign=c, width=0.2\textwidth]{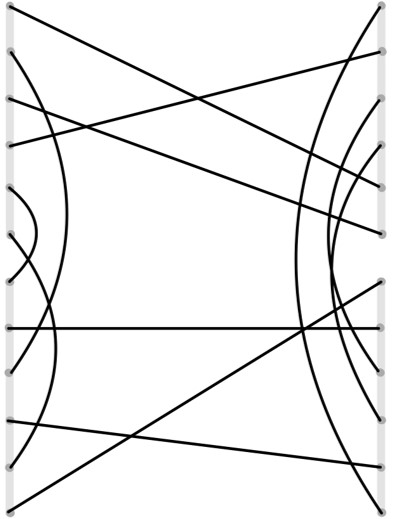}
\end{equation}
Another example relates to the calculation of the 2-replica case in the previous section~\ref{section:1/2 replica}: $B_{0}^{(2)}=A_5+A_7$, $B_{1}^{(2)}=A_2+A_4+A_8$, and $B_{2}^{(2)}=A_1+A_3+A_6$, where $A_{i}$ is defined in equation~(\ref{eq:definition of A_i}).

Using equation~(\ref{eq:matrix repre in 2-replica}), we find that $\mathcal{L}_2B_0^{(2)}=0$, $\mathcal{L}_2B_1^{(2)}=-B_1^{(2)}+4N^{-1}B_0^{(2)}$, and $\mathcal{L}_2B_2^{(2)}=-(2+2N^{-1})B_2^{(2)}+N^{-1}B_1^{(2)}$. We see that the action of $\mathcal{L}_2$ is closed in these three bases, which means the vector space defined by $\text{Span}\{B_0^{(2)},B_1^{(2)},B_2^{(2)}\}$ forms a three-dimensional representation of $\mathcal{L}_2$.

Motivated by this special result at $k=2$, in this appendix, we show that the linear space $\text{Span}\{B_{p}^{(k)} | p=0,1,\cdots,k\}$ forms a $(k+1)$-dimensional representation of $\mathcal{L}_k$. We further show that the representation matrix of $\mathcal{L}_k$ is upper-triangular. Therefore, we can determine the eigenvalues from the diagonal entries of the representation matrix, which turn out to be $\{-\left[p+N^{-1}p(p-1)\right] | p=0,1,\cdots,k\}$.

Let's act $\mathcal{L}_k$ on $B_p^{(k)}$. The analysis is similar to that in the second excited states, retaining the four terms recorded in table~\ref{table:subset of high excited states}.

Therefore, we see that the action of $\mathcal{L}_{k}$ can either map diagrams in $B_p^{(k)}$ to diagrams in $B_{p-1}^{(k)}$ (process $\boxed{1'''}$), or map diagrams in $B_p^{(k)}$ to diagrams in $B_{p}^{(k)}$ (processes $\boxed{2'''}, \boxed{3'''}, \boxed{4'''}$). By the permutation symmetry of $\mathcal{L}_k$ and $B_p^{(k)}, B_{p-1}^{(k)}$, we conclude that $\mathcal{L}_k$ takes $B_p^{(k)}$ to be a linear superposition of $B_p^{(k)}$ and $B_{p-1}^{(k)}$:
\begin{equation}
\begin{aligned}
\mathcal{L}_kB_{p}^{(k)} &= -kB_p^{(k)} + N^{-1}\left(\boxed{1'''}+\boxed{2'''}-\boxed{3'''}-\boxed{4'''}\right) B_p^{(k)}\\
&= -kB_p^{(k)} + N^{-1}p^2 \frac{\#\big(B_p^{(k)}\big)}{\#\big(B_{p-1}^{(k)}\big)} \cdot B_{p-1}^{(k)} + N^{-1}(k-p)N \cdot B_p^{(k)} - 2 \times N^{-1} \binom{p}{2} \cdot B_p^{(k)}\\
&= -\left[p + N^{-1}p(p-1)\right] \cdot B_p^{(k)} + N^{-1}(k-p+1)^2 \cdot B_{p-1}^{(k)}, \ B_{-1}^{(k)} \equiv 0
\end{aligned}
\end{equation}
This shows that the representation matrix of $\mathcal{L}_k$ is upper-triangular, and therefore we can determine the eigenvalues from the diagonal entries, as claimed at the beginning of this appendix. We also note that when taking $k=2$, these equations successfully reproduce the two-replica results.

\bibliographystyle{JHEP}
\bibliography{biblio}

\end{document}